\documentclass[twocolumn, letter, superscriptaddress]{revtex4-1}

\usepackage{graphicx}
\usepackage{hyperref}
\usepackage{bm}
\usepackage{color}
\usepackage{enumitem} 
\usepackage[T1]{fontenc}
\usepackage{lipsum}

\usepackage{amsmath}
\usepackage{physics}
\usepackage{amsmath, array} 
\usepackage[normalem]{ulem}
\usepackage{soul,xcolor}
\usepackage{physics}
\usepackage{lineno}

\begin{document}

\title{Possible many-body localization in a long-lived finite-temperature ultracold quasi-neutral molecular plasma}

\author{John Sous}  
\affiliation{Department of Physics \& Astronomy, University of British Columbia, Vancouver, BC V6T 1Z3, Canada}
\affiliation{Stewart Blusson Quantum Matter Institute, University of British Columbia, Vancouver, British Columbia, V6T 1Z4, Canada}
\author{Edward Grant}
\email[Author to whom correspondence should be addressed. Electronic mail:  ]
{edgrant@chem.ubc.ca}
\affiliation{Department of Physics \& Astronomy, University of British Columbia, Vancouver, BC V6T 1Z3, Canada}
\affiliation{Department of Chemistry, University of British Columbia, Vancouver, BC V6T 1Z3, Canada}

\begin{abstract}
We argue that the quenched ultracold plasma presents an experimental platform for studying quantum many-body physics of disordered systems in the long-time and finite energy-density limits. We consider an experiment that quenches a plasma of nitric oxide to an ultracold system of Rydberg molecules, ions and electrons that exhibits a long-lived state of arrested relaxation.  The qualitative features of this state fail to conform with classical models.  Here, we develop a microscopic quantum description for the arrested phase based on an effective many-body spin Hamiltonian that includes both dipole-dipole and van der Waals interactions.  This effective model appears to offer a way to envision the essential quantum disordered non-equilibrium physics of this system.

\end{abstract}

\maketitle

{\em Introduction.}--- Quantum mechanics serves well to describe the discrete low-energy dynamics of isolated microscopic many-body systems \cite{Sakurai}.  The macroscopic world conforms with the laws of Newtonian mechanics \cite{Goldstein}.  Quantum statistical mechanics \cite{Kardar2} bridges these realms by treating the quantum mechanical properties of an ensemble of particles statistically and characterizing the properties of the system in terms of state properties (temperature, chemical potential, etc.), in an approach that implies a complex phase space of trajectories with ergodic dynamics \cite{ChoasRev}.  However, this is not always the case, and the macroscopic description of quantum many-body systems that fail to behave as expected statistically remains today as a key unsolved problem \cite{NonEqRev, nandkishoreRev}.

Ergodicity, when present in an isolated quantum many-body system, emerges as the system thermalizes in a unitary evolution that spreads information among all the subspaces of the system.  The subspaces act as thermal reservoirs for each other.  Most known many-body systems thermalize in this fashion, obeying the Eigenstate Thermalization Hypothesis (ETH) \cite{nandkishoreRev,ChoasRev, Deustch, Tasaki, Srednicki, Rigol_Olshanii} which holds that the eigenstates of corresponding many-body Hamiltonians are thermal.

Exceptions include fine-tuned integrable systems \cite{Rigol_integrable}, and the class of so-called many-body localized (MBL) systems \cite{BAA,nandkishoreRev}, which have attracted intense interest in recent years.  Such systems do not thermalize at finite energy densities and are therefore non-ergodic.  Disorder in a landscape of interactions preserves memory of the initial local conditions for infinitely long times.  MBL phases cannot be understood in terms of conventional quantum statistical mechanics  \cite{Altman_MBLtransition, Khemani_MBLtransition}.

MBL has been observed in deliberately engineered experimental systems with ultracold atoms in one and two-dimensional optical lattices \cite{MBL_Hubbard, MBL1D, Monroe_MBL, MBL2D_2,MBLopensystem, MBL2D_1}. In such cases, tuning of the lattice parameters allows investigation of the phase diagram of the system as a function of disorder strength.  However, such ultracold systems suffer from decoherence, confining localization to short timescales and low energy densities. 

It is important to determine experimentally whether conditions exist under which MBL can persist for long times at finite temperatures, and to understand if such a robust macroscopic quantum many-body state can occur naturally in an interacting quantum system without deliberate tuning of experimental parameters.  Such a realization could pave the way to exotic quantum effects, such as entangled macroscopic objects and localization-protected quantum order \cite{Loc_quantumorder,Moore}, which could have societal and technological implications \cite{MBLmobile}.  

Motivated by these questions, we have explored the quenched ultracold molecular plasma as an arena in which to study quantum many-body effects in the long-time and finite energy-density limits \cite{Morrison2008,Haenel}. The ultracold plasma system offers complexity, as encountered in quantum materials, but evolves from state-selected initial conditions that allow for a description in terms of a specific set of atomic and molecular degrees of freedom.  

Experimental work has recently established laboratory conditions under which a high-density molecular ultracold plasma evolves from a cold Rydberg gas of nitric oxide, adiabatically sequesters energy in a reservoir of global mass transport, and relaxes to form a spatially correlated, strongly coupled plasma \cite{SchulzWeiling,Haenel}.  This system naturally evolves to form an arrested phase that has a long lifetime with respect to recombination and neutral dissociation, and a very slow rate of free expansion. These volumes exhibit state properties that are independent of initial quantum state and density, parameters which critically affect the timescale of relaxation, suggesting a robust process of self-assembly that reaches an arrested state, far from conventional thermal equilibrium. 

Departure from classical models suggests localization in the disposition of energy \cite{Haenel}.  In an effort to explain this state of arrested relaxation, we have developed a quantum mechanical description of the system in terms of power law interacting spin model, which allows for the possibility of slow dynamics or MBL

{\em Experiment.}--- Double-resonant pulsed-laser excitation of nitric oxide entrained in a supersonic molecular beam forms a characteristic Gaussian ellipsoid volume of state-selected Rydberg gas that propagates in $z$ with a well-defined velocity, longitudinal temperature ($T_{||}=500$ mK) transverse temperature ($T_{\perp}<5$ mK) and precisely known initial density in a range from $\rho_0=10^{10}$ to $10^{12}~{\rm cm^{-3}}$ (See Figure \ref{fig:exp} and References \cite{MSW_tutorial, [{See Supplemental Material at [URL will be inserted by publisher] for details including further references \cite{Gallagher,Fielding,Mansbach,PVS,Bixon,GallagherNO,Chupka:1993,Saquet2011,Saquet2012,Dorozhkina,Sadeghi:2012,agranovich,BrownC,Pillet_Cs,Zoubi1,Imbrie,Sarang1, Sarang2,RareRegions_rev, Thermal_inclusions}}] Supp}).  

Rydberg molecules in the leading edge of the nearest-neighbour distance distribution interact to produce NO$^+$ ions and free electrons \cite{Sadeghi:2014}.  Electron-Rydberg collisions trigger an ionization avalanche on a time scale from nanoseconds to microseconds depending on initial density and principal quantum number, $n_0$.  

Inelastic collisions heat electrons and the system proceeds to a quasi-equilibrium of ions, electrons and high-Rydberg molecules of nitric oxide.  This relaxation and the transient state it produces entirely parallels the many observations of ultracold plasma evolution in atomic systems under the conditions of a magneto-optical trap (MOT) \cite{WalzFlannigan}.  

We see this avalanche unfold directly in sequences of density-classified selective field ionization spectra measured as a function of delay after initial formation of the Rydberg gas \cite{Haenel}.  For a moderate $\rho_0 = 3 \times 10^{11}~{\rm cm^{-3}}$, the ramp-field signal of the selected Rydberg state, $n_0$ gives way on a 100 ns timescale to form the selective field ionization (SFI) spectrum of a system in which electrons bind very weakly to single ions in a narrow distribution of high Rydberg states or in a quasi-free state held by the plasma space charge \cite{[{See Supplemental Material at [URL will be inserted by publisher] for details including further references \cite{Gallagher,Fielding,Mansbach,PVS,Bixon,GallagherNO,Chupka:1993,Saquet2011,Saquet2012,Dorozhkina,Sadeghi:2012,agranovich,BrownC,Pillet_Cs,Zoubi1,Imbrie,Sarang1, Sarang2,RareRegions_rev, Thermal_inclusions}}] Supp}.  

The peak density of the plasma decays for as much as 10 $\mu$s until it reaches a value of $\sim 4 \times 10^{10}~{\rm cm^{-3}}$, independent of the initially selected $n_0$ and $\rho_0$.  Thereafter the number of charged particles remains constant for at least a millisecond.  On this hydrodynamic timescale, the plasma bifurcates, disposing substantial energy in the relative velocity of plasma volumes separating in $\pm x$, the cross-beam axis of laser propagation \cite{SchulzWeiling}.  

\begin{figure}[t]
    \centering
        \includegraphics[width=1.0\columnwidth]{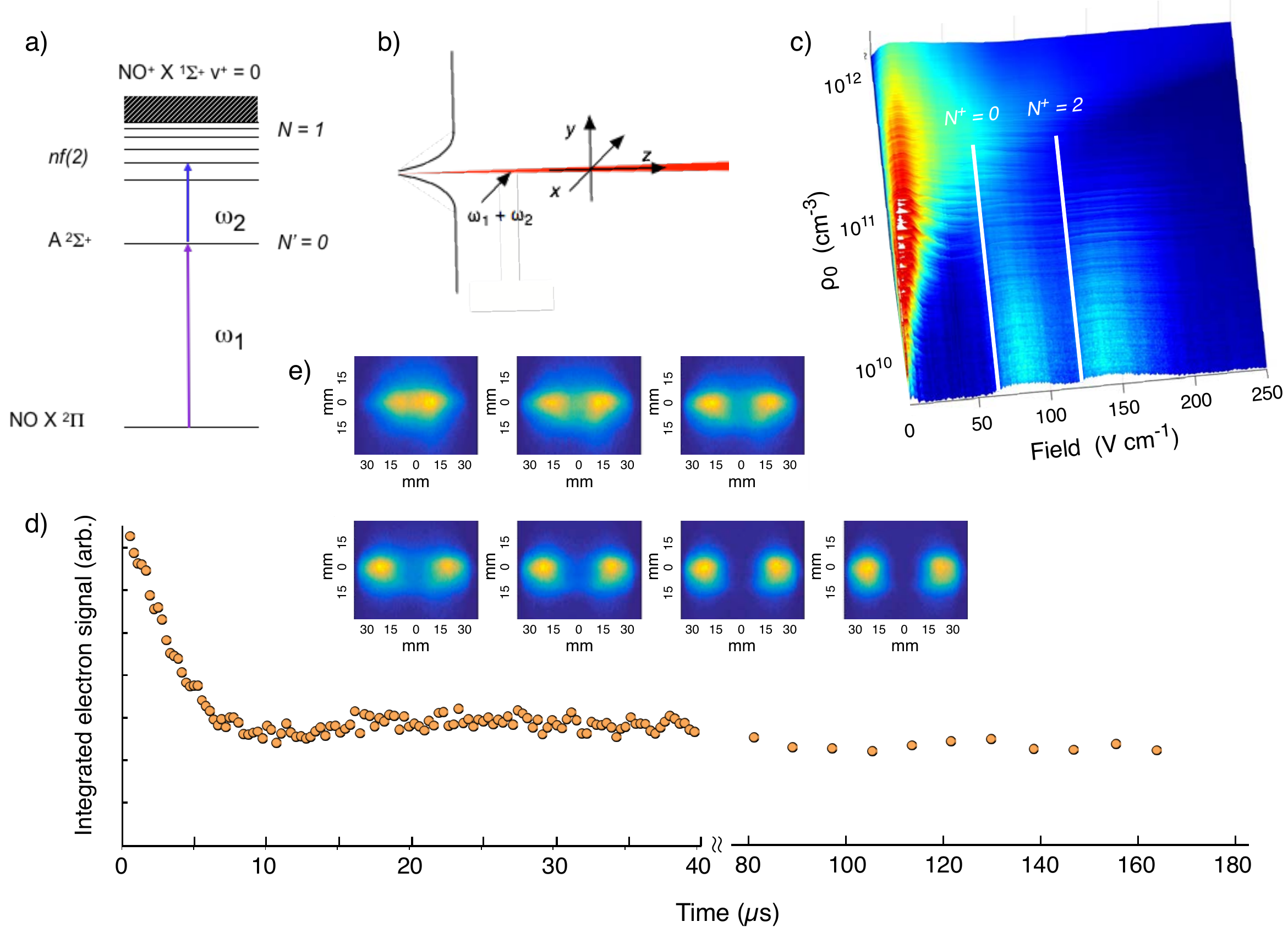}
    \caption{a) Double-resonant selection of the initial quantum state of the $n_0f(2)$ Rydberg gas. b) Laser-crossed differentially pumped supersonic molecular beam.  c) Selective field ionization spectrum after 500 ns evolution, showing the signal of weakly bound electrons combined with a residual population of $49f(2)$ Rydberg molecules.  After 10 $\mu$s, this population sharpens to signal only high-$n$ Rydbergs and plasma electrons.  d)  Integrated electron signal as a function of evolution time from 0 to 160 $\mu$s.  Note the onset of the arrest phase before 10 $\mu$s.  Timescale compressed by a factor of two after 80 $\mu$s.  e)  $x,y$-integrated images recorded after a flight time of 400 $\mu$s with $n_0=40$ for initial Rydberg gas peak densities varying from $2 \times 10^{11}$ to $1 \times 10^{12}~{\rm cm}^{-3}$.  All of these images exhibit the same peak density, ~$1 \times 10^7~{\rm cm}^{-3}$. }
    \label{fig:exp}
\end{figure}

The avalanche to plasma proceeds at a rate predicted with accuracy by semi-classical coupled rate equations  \cite{Haenel, [{See Supplemental Material at [URL will be inserted by publisher] for details including further references \cite{Gallagher,Fielding,Mansbach,PVS,Bixon,GallagherNO,Chupka:1993,Saquet2011,Saquet2012,Dorozhkina,Sadeghi:2012,agranovich,BrownC,Pillet_Cs,Zoubi1,Imbrie,Sarang1, Sarang2,RareRegions_rev, Thermal_inclusions}}] Supp}.  This picture also calls for the rapid collisional relaxation of Rydberg molecules, accompanied by an increase in electron temperature to 60 K or more.  Bifurcation accounts for a loss of electron energy.  But, the volumes that remain cease to evolve, quenching instead to form an arrested phase that expands slowly, at a rate reflecting an initial electron temperature no higher than a few degrees Kelvin.  These volumes show no sign of loss owing to the fast dissociative recombination of NO$^+$ ions with electrons predicted classically for low $T_e$ \cite{Schneider}, or predissociation of NO Rydbergs, well-known to occur with relaxation in $n$ \cite{Remacle}.

Thus, from the experiment, we learn that 5 $\mu$s after avalanche begins, Rydberg relaxation ceases.  We detect no sign of ion acceleration by hot electrons and the surviving number of ions and electrons remains constant for the entire remaining observation period, extending to as long as 1 ms. With the vast phase space available to energized electrons and neutral nitrogen and oxygen atom fragments, this persistent localization of energy in the electrostatic separation of cold ions and electrons represents a very significant departure from a thermalized phase.  Current experimental evidence thus points strongly to energy localization and absence of thermalization within the accessible time of the experiment.

{\em Molecular physics of the arrested phase.}---
Direct measurements of its electron binding energy together with its observed expansion rate establish experimentally that the bifurcated plasma contains only high-Rydberg molecules ($n>80$) and NO$^+$ ions in combination with cold electrons (initial $T_e < 5$ K) bound by the space charge.  As noted above, semi-classical models mixing these species in any proportion predict thermal relaxation, electron heating, expansion and dissipation on a rapid timescale with very evident consequences completely unobserved in the experiment.  Instead, beyond an evolution time of 10 $\mu$s or less, we find that the plasma settles in a state of arrested relaxation of canonical density and low internal energy manifested by a slow free expansion.  

\begin{figure}[t]
    \centering
        \includegraphics[width=.7 \columnwidth]{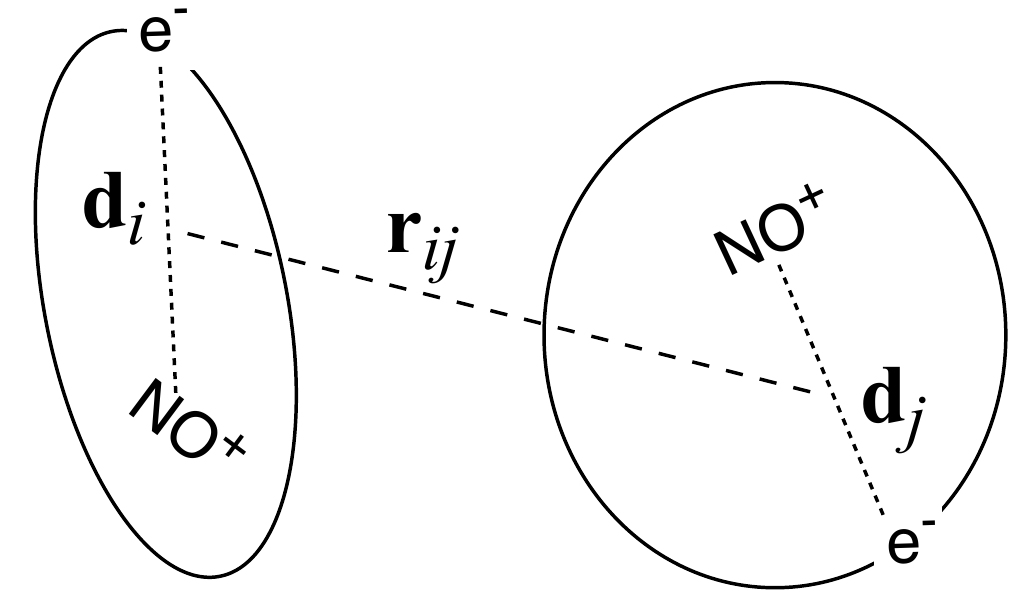}
    \caption{Schematic representation of NO$^+$ core ions, paired with extravalent electrons to form interacting dipoles {\bf d}$_{i}$ and {\bf d}$_{j}$, separated by ${\bf r}_{ij} = {\bf r}_i - {\bf r}_j$. }
    \label{fig:dipole} 
\end{figure}

To describe this apparent state of suppressed relaxation, we proceed now to develop a formal representation of the predominant interactions in this arrested phase.  Under the evidently cold, quasi-neutral conditions of the relaxed plasma, ions pair with extravalent electrons to form dipoles which interact as represented schematically in Figure \ref{fig:dipole}.   

Assuming an intermolecular spacing that exceeds the dimensions of individual ion-electron separations, we can describe the Coulomb interactions represented in Figure \ref{fig:dipole} in terms of a simple Hamiltonian:
\begin{equation}
{H} = \sum_i \left ({\frac{{\bf P}_i^2}{2m}} + h_i \right) + \sum_{i,j}{{{{V}}}_{ij}}
\label{equ:ham}
\end{equation}
where $h_i$ describes the local relationship of each electron with its proximal NO$^+$ core.  This local representation extends to account for the interactions of a bound extravalent electron with vibrational, rotational and electronic degrees of freedom of the core, as described, for example, by Multichannel Quantum Defect Theory \cite{Greene}.  Each ion-electron pair has momentum, ${{\bf P}_i}$ and ${{{{V}}}_{ij}} \equiv {{{{V}}}}({\bf r_i - r_j})$ describes the potential energy of the interacting multipoles, represented in Figure \ref{fig:dipole} to lowest order as induced dipoles with an interaction defined by  ${{V}^{\rm dd}_{ij}} = \left [{\bf d}_i \cdot  {\bf d}_j - 3({\bf d}_i\cdot {\bf r}_{ij})({\bf d}_j \cdot {\bf r}_{ij})\right ]  / {{ {\bf r}_{ij}^3}}$, where for simplicity we average over the anisotropy of the dipole-dipole interaction.  

The plasma also very likely includes ion-electron pairs of positive total energy.  This implies the existence of local Hamiltonians of much greater complexity that define quasi-Rydberg bound states with dipole and higher-order moments formed by the interaction of an extravalent electron with more than one ion.

Representing the eigenstates of $h_i$ by $\ket{e_i}$, we can write a reduced Hamiltonian for the pairwise dipole-dipole interactions \cite{Pupillo, Krems_book1} in the arrested phase:
\begin{equation}
H^{\rm dd} =  \sum_i  {\frac{{\bf P}_i^2}{2m}} + \sum_{i,j}{{V}^{\rm dd}_{ij}}
\label{equ:dham}
\end{equation}
where we evaluate ${{V}^{\rm dd}_{ij}}$ in the $\ket{e_i}$ basis.  

Note that such a Hamiltonian usually refers to the case where a narrow bandwidth laser prepares a Rydberg gas in which a particular set of dipole-dipole interactions give rise to a small, specific set of coupled states \cite{Pillet,Low,Firstenberg}.   By contrast, the molecular ultracold plasma forms spontaneously by processes of avalanche and quench to populate a great many different states that evolve spatially without the requirement of light-matter coherence or reference to a dipole blockade of any kind.  

This system relaxes to a quenched regime of ultracold temperature, from which it expands radially at a rate of a few meters per second.  Dipolar energy interactions proceed on a much faster timescale \cite{rydbergs1,rydbergs5,Barredo,rydberg_Gross}.  Cross sections for close-coupled collisions are minuscule by comparison \cite{Wester}.  We can thus assume that the coupled states defined by dipole-dipole interactions evolve adiabatically with the motion of ion centres.  
 
This separation of timescales enables us to write an effective Hamiltonian describing pairwise interactions that slowly evolve in an instantaneous frame of slowly moving ions and Rydberg molecules: $H_{\rm eff} = \mathcal{P} \sum_{i,j}{{V}^{dd}_{ij}}$, where $\mathcal{P}$ represents a projector onto the low-energy degrees of freedom owing to dipole-dipole coupling.

{\em Effective many-body Hamiltonian.}--- Considering pairwise dipolar interactions between ion-electron pairs, we choose a set of basis states $\ket{e^1}$, $\ket{e^2}$, $ ... $ $\ket{e^L}$ that  spans the low-energy regime.  The superscript with lower (higher) integer label refers to the state with larger (smaller) electron binding energy. 

Quenching gives rise to a vast distribution of rare resonant pair-wise interactions, creating a random potential landscape.  Dipole-dipole interactions in this dense manifold of basis states cause excitation exchange. In the disorder potential, these processes are dominated by low energy-excitations involving $L$ states in number, where we expect $L$ to be small (from 2 to 4).  The most probable interactions select $L$-level systems composed of {\it different } basis states from dipole to dipole.  

In a limit of dipole-dipole coupling, we can represent pairwise excitations by spins with energies, $\epsilon_i$, and exchange interactions governed by an XY model Hamiltonian \cite{Sachdevbook, [{See Supplemental Material at [URL will be inserted by publisher] for details including further references \cite{Gallagher,Fielding,Mansbach,PVS,Bixon,GallagherNO,Chupka:1993,Saquet2011,Saquet2012,Dorozhkina,Sadeghi:2012,agranovich,BrownC,Pillet_Cs,Zoubi1,Imbrie,Sarang1, Sarang2,RareRegions_rev, Thermal_inclusions}}] Supp} that describes these interactions in terms of their effective spin dynamics:
\begin{eqnarray} \label{XY}
&H_{\rm eff}& =  \sum_i \epsilon_i \hat{S}^z_i + \sum_{i,j} J_{ij} (\hat{S}^+_i \hat{S}^-_j + h.c.)
\label{eqn:XY}
\end{eqnarray}
where $\hat{S}$ in each case denotes a spin-$L$ operator defined as $\hat{S}^\gamma = \hbar \hat{\sigma}^\gamma /2 $, for which $\sigma^\gamma$ is the corresponding spin-$L$ Pauli matrix that spans the space of the $L$ active levels and $\gamma = x,y$ or $z$. $h.c.$ refers to Hermitian conjugate.

This Hamiltonian reflects both the diagonal and off-diagonal disorder created by the variation in $L$-level system from dipole to dipole.  The first term in $H_{\rm eff}$ describes the diagonal disorder arising from random contributions to the on-site energy of any particular dipole owing to its random local environment. In spin language, $\sum_i \epsilon_i \hat{S}^z_i$ represents a Gaussian-distributed random local field of width $W$.  The representative SFI spectrum in Figure \ref{fig:exp} directly gauges a $W$ of $\sim 500$ GHz for the quenched ultracold plasma.

In the second term, $J_{ij} = {t_{ij}}/{ {r}_{ij}^3}$ determines the off-diagonal disordered amplitudes of the spin flip-flops.  To visualize the associated disorder, recognize that the second term varies as $t_{ij} \propto \abs{\textbf{d}_i} \abs{\textbf{d}_j}$, where every interaction selects a different ${\bf d}_i$ and ${\bf d}_j$.  Over the present range of $W$, a simple pair of dipoles formed by $s$ and $p$ Rydberg states of the same $n$ couple with a $t_{ij}$ of 75 GHz $\mu$m$^3$ \cite{Zoubi2}.  Note that $t_{ij}$ falls exponentially with the difference in principal quantum numbers, $\Delta n_{ij}$ \cite{Samboy}.

{\em Induced Ising interactions.} 
In the limit $|J_{ij}| << W$ most appropriate to the experiment, sequences of interactions can add an Ising term that describes a van der Waals shift of pairs of dipoles  \cite{[{See Supplemental Material at [URL will be inserted by publisher] for details including further references \cite{Gallagher,Fielding,Mansbach,PVS,Bixon,GallagherNO,Chupka:1993,Saquet2011,Saquet2012,Dorozhkina,Sadeghi:2012,agranovich,BrownC,Pillet_Cs,Zoubi1,Imbrie,Sarang1, Sarang2,RareRegions_rev, Thermal_inclusions}}] Supp,Burin1}.   These processes occur with an amplitude, $U_{ij} \approx {J_{ij}^2\widetilde{J}}/{W^2}$, where $\widetilde{J}$  estimates $J_{ij}$, for an average value of $t_{ij}$ at an average distance separating spins.   $U_{ij}$ is inherently random owing to the randomness in $J_{ij}$.  

Together, these results lead us to a general spin model with dipole-dipole and van der Waals interactions \cite{[{See Supplemental Material at [URL will be inserted by publisher] for details including further references \cite{Gallagher,Fielding,Mansbach,PVS,Bixon,GallagherNO,Chupka:1993,Saquet2011,Saquet2012,Dorozhkina,Sadeghi:2012,agranovich,BrownC,Pillet_Cs,Zoubi1,Imbrie,Sarang1, Sarang2,RareRegions_rev, Thermal_inclusions}}] Supp}:
\begin{eqnarray} \label{XY-Ising}
H_{\rm eff}& = & \sum_i \epsilon_i \hat{S}^z_i + \sum_{i,j} J_{ij} (\hat{S}^+_i \hat{S}^-_j + h.c.) \nonumber \\
&+& \sum_{i,j} U_{ij} \hat{S}^z_i \hat{S}^z_j 
\end{eqnarray}
where $U_{ij} = {D_{ij}}/{r}_{ij}^6$ and $D_{ij} = {t_{ij}^2\widetilde{J}}/{W^2}$. 

{\em Discussion:  Localization versus glassy behaviour and slow dynamics.}--- The complexity of this Hamiltonian places an exact solution of Eq (\ref{XY-Ising}) beyond reach for the conditions of the plasma.  But, we can gauge some likely properties of such a solution by analogy to published work on simpler systems.  

In the single-spin limit, this Hamiltonian reduces to the dipolar XY model, which has been studied by locator expansion methods measuring the probability of resonant pairs \cite{Anderson, Levitov}.  When $J_{ij}$ scales by a power law $\alpha$ that equals the dimension $d$, a single-spin model with diagonal disorder displays critical behaviour characterized by extended states with subdiffusive dynamics \cite{Anderson, Levitov}.  
Dipolar systems in three dimensions can form extended states, but yet exhibit non-ergodic behaviour  \cite{Deng_nonergodic}.

Off-diagonal disorder in the presence of long-range spin flip-flop interactions of arbitrary order in one dimension yields algebraic localization as opposed to exponential Anderson localization, challenging the generality of the rule that says systems must delocalize for $\alpha \leq d$ \cite{Deng_algebraic}.

The many-body problem is more involved, because the van der Waals term forms off-diagonal matrix elements in the resonant pair states \cite{Burin2}. This mechanism couples distant resonant pairs, transferring energy from one pair to the other to cause delocalization.  A study of power-law coupled systems predicts that spin flip-flops (order $\alpha$) and spin Ising interactions (order $\beta$) in an iterated pairs configuration in which $\beta \leq \alpha$ will localize for $\beta/2 > d$ \cite{Yao}.  

A locator expansion approach developed for $\beta > \alpha$ applied to Eq (\ref{XY-Ising}) confined to diagonal disorder predicts a critical dimension, $d_c = 2$ \cite{Burin1}.  For the case of $d>d_c$, this theory holds that a diverging number of resonances drives delocalization whenever the number of dipoles exceeds a critical value $N_c$.  

A system described by Eq \ref{XY-Ising} for the conditions under which we observe arrest requires a number of dipoles, $N_c=(W/\widetilde{J})^4 \approx 3 \times 10^9$ to delocalize \cite{[{See Supplemental Material at [URL will be inserted by publisher] for details including further references \cite{Gallagher,Fielding,Mansbach,PVS,Bixon,GallagherNO,Chupka:1993,Saquet2011,Saquet2012,Dorozhkina,Sadeghi:2012,agranovich,BrownC,Pillet_Cs,Zoubi1,Imbrie,Sarang1, Sarang2,RareRegions_rev, Thermal_inclusions}}] Supp}.  This theoretical threshold deemed necessary for resonance delocalization exceeds the number of molecules found experimentally in the quenched ultracold plasma by more than an order of magnitude \cite{[{See Supplemental Material at [URL will be inserted by publisher] for details including further references \cite{Gallagher,Fielding,Mansbach,PVS,Bixon,GallagherNO,Chupka:1993,Saquet2011,Saquet2012,Dorozhkina,Sadeghi:2012,agranovich,BrownC,Pillet_Cs,Zoubi1,Imbrie,Sarang1, Sarang2,RareRegions_rev, Thermal_inclusions}}] Supp}.  

Moreover, as Nandkishore and Sondhi have pointed out \cite{Sondhi}, locator expansion arguments might not hold generally, and low-order power law interactions could well give rise to  MBL in higher dimensions.  Their arguments build on the idea that, in many systems, long-range interactions can drive a system to form correlated phases in which emergent short-range interactions can be well characterized by a locator expansion perturbation theory approach. In this context, MBL with long-range interactions in higher dimensions becomes quite possible.

A related study has investigated the behavior of a three-dimensional dipolar system of nitrogen-vacancy color centers in diamond in the presence of onsite disorder \cite{Lukin_critical}. The experimental results point to slow dynamics consistent with our observations.  

The forgoing analysis suggests that the model defined by Eq (\ref{XY-Ising}) ought to exhibit some form of localization or at least very slow dynamics, since all the terms in the Hamiltonian are disordered and the terms responsible for delocalization ($J_{ij}$ and $U_{ij}$) are expected to be much smaller than $W$.  This seems to be what we see in the experiment

{\em Concluding remarks.}---  This work has argued that the quenched ultracold plasma forms an arrested phase possibly governed by quantum disordered non-equilibrium physics in long-time and finite energy-density limits.  In an effort to support this notion, we have suggested that the evident and certainly present quantum dipolar interactions can be usefully described by a disordered spin model and analyzed its properties in the limit of strong onsite disorder by analogy with theoretical results for simpler dipolar systems.  

Considering the challenge of scale confronting the accurate numerical solution of large disordered problems and the apparent contradiction of available theoretical results \cite{Scardicchio_allD, Pollet_MBL, Roeck_griffith, Scardicchio_ergodicity}, experimental systems stand to play an important role in understanding localization and slow dynamics.  The results presented here call in particular for further experimental and theoretical efforts to probe the physics of localization in long-range interacting systems of higher dimension.  The quenched ultracold plasma appears to offer a view of large-scale quantum dynamics in a regime inaccessible to optical lattices and solid-state materials.  

\begin{acknowledgements}
 
This work was supported by the US Air Force Office of Scientific Research (Grant No. FA9550-17-1-0343), together with the Natural Sciences and Engineering research Council of Canada (NSERC) and the Stewart Blusson Quantum Matter Institute (SBQMI).  JS gratefully acknowledges support from the Harvard-Smithsonian Institute for Theoretical Atomic, Molecular and Optical Physics (ITAMP).  We have benefited from helpful interactions with Rahul Nandkishore, Shivaji Sondhi and Alexander Burin.  We also appreciate discussions with Joshua Cantin and Roman Krems.  

\end{acknowledgements}

\bibliography{Rydberg_localization}

\begin{thebibliography}{79}%
\makeatletter
\providecommand \@ifxundefined [1]{%
 \@ifx{#1\undefined}
}%
\providecommand \@ifnum [1]{%
 \ifnum #1\expandafter \@firstoftwo
 \else \expandafter \@secondoftwo
 \fi
}%
\providecommand \@ifx [1]{%
 \ifx #1\expandafter \@firstoftwo
 \else \expandafter \@secondoftwo
 \fi
}%
\providecommand \natexlab [1]{#1}%
\providecommand \enquote  [1]{``#1''}%
\providecommand \bibnamefont  [1]{#1}%
\providecommand \bibfnamefont [1]{#1}%
\providecommand \citenamefont [1]{#1}%
\providecommand \href@noop [0]{\@secondoftwo}%
\providecommand \href [0]{\begingroup \@sanitize@url \@href}%
\providecommand \@href[1]{\@@startlink{#1}\@@href}%
\providecommand \@@href[1]{\endgroup#1\@@endlink}%
\providecommand \@sanitize@url [0]{\catcode `\\12\catcode `\$12\catcode
  `\&12\catcode `\#12\catcode `\^12\catcode `\_12\catcode `\%12\relax}%
\providecommand \@@startlink[1]{}%
\providecommand \@@endlink[0]{}%
\providecommand \url  [0]{\begingroup\@sanitize@url \@url }%
\providecommand \@url [1]{\endgroup\@href {#1}{\urlprefix }}%
\providecommand \urlprefix  [0]{URL }%
\providecommand \Eprint [0]{\href }%
\providecommand \doibase [0]{http://dx.doi.org/}%
\providecommand \selectlanguage [0]{\@gobble}%
\providecommand \bibinfo  [0]{\@secondoftwo}%
\providecommand \bibfield  [0]{\@secondoftwo}%
\providecommand \translation [1]{[#1]}%
\providecommand \BibitemOpen [0]{}%
\providecommand \bibitemStop [0]{}%
\providecommand \bibitemNoStop [0]{.\EOS\space}%
\providecommand \EOS [0]{\spacefactor3000\relax}%
\providecommand \BibitemShut  [1]{\csname bibitem#1\endcsname}%
\let\auto@bib@innerbib\@empty
\bibitem [{\citenamefont {Sakurai}(2014)}]{Sakurai}%
  \BibitemOpen
  \bibfield  {author} {\bibinfo {author} {\bibfnamefont {J.~J.}\ \bibnamefont
  {Sakurai}},\ }\href@noop {} {\emph {\bibinfo {title} {Modern Quantum
  Mechanics}}}\ (\bibinfo  {publisher} {Pearson (London)},\ \bibinfo {year}
  {2014})\BibitemShut {NoStop}%
\bibitem [{\citenamefont {Goldstein}(2011)}]{Goldstein}%
  \BibitemOpen
  \bibfield  {author} {\bibinfo {author} {\bibfnamefont {H.}~\bibnamefont
  {Goldstein}},\ }\href@noop {} {\emph {\bibinfo {title} {Classical
  mechanics}}}\ (\bibinfo  {publisher} {Pearson Education India},\ \bibinfo
  {year} {2011})\BibitemShut {NoStop}%
\bibitem [{\citenamefont {Kardar}(2007)}]{Kardar2}%
  \BibitemOpen
  \bibfield  {author} {\bibinfo {author} {\bibfnamefont {M.}~\bibnamefont
  {Kardar}},\ }\href@noop {} {\emph {\bibinfo {title} {Statistical physics of
  particles}}}\ (\bibinfo  {publisher} {Cambridge University Press},\ \bibinfo
  {year} {2007})\BibitemShut {NoStop}%
\bibitem [{\citenamefont {D'Alessio}\ \emph {et~al.}(2016)\citenamefont
  {D'Alessio}, \citenamefont {Kafri}, \citenamefont {Polkovnikov},\ and\
  \citenamefont {Rigol}}]{ChoasRev}%
  \BibitemOpen
  \bibfield  {author} {\bibinfo {author} {\bibfnamefont {L.}~\bibnamefont
  {D'Alessio}}, \bibinfo {author} {\bibfnamefont {Y.}~\bibnamefont {Kafri}},
  \bibinfo {author} {\bibfnamefont {A.}~\bibnamefont {Polkovnikov}}, \ and\
  \bibinfo {author} {\bibfnamefont {M.}~\bibnamefont {Rigol}},\ }\href@noop {}
  {\bibfield  {journal} {\bibinfo  {journal} {Adv. Phys.}\ }\textbf {\bibinfo
  {volume} {65}},\ \bibinfo {pages} {239} (\bibinfo {year} {2016})}\BibitemShut
  {NoStop}%
\bibitem [{\citenamefont {Polkovnikov}\ \emph {et~al.}(2011)\citenamefont
  {Polkovnikov}, \citenamefont {Sengupta}, \citenamefont {Silva},\ and\
  \citenamefont {Vengalattore}}]{NonEqRev}%
  \BibitemOpen
  \bibfield  {author} {\bibinfo {author} {\bibfnamefont {A.}~\bibnamefont
  {Polkovnikov}}, \bibinfo {author} {\bibfnamefont {K.}~\bibnamefont
  {Sengupta}}, \bibinfo {author} {\bibfnamefont {A.}~\bibnamefont {Silva}}, \
  and\ \bibinfo {author} {\bibfnamefont {M.}~\bibnamefont {Vengalattore}},\
  }\href {\doibase 10.1103/RevModPhys.83.863} {\bibfield  {journal} {\bibinfo
  {journal} {Rev. Mod. Phys.}\ }\textbf {\bibinfo {volume} {83}},\ \bibinfo
  {pages} {863} (\bibinfo {year} {2011})}\BibitemShut {NoStop}%
\bibitem [{\citenamefont {Nandkishore}\ and\ \citenamefont
  {Huse}(2015)}]{nandkishoreRev}%
  \BibitemOpen
  \bibfield  {author} {\bibinfo {author} {\bibfnamefont {R.}~\bibnamefont
  {Nandkishore}}\ and\ \bibinfo {author} {\bibfnamefont {D.~A.}\ \bibnamefont
  {Huse}},\ }\href@noop {} {\bibfield  {journal} {\bibinfo  {journal} {Annu.
  Rev. Condens. Matter Phys.}\ }\textbf {\bibinfo {volume} {6}},\ \bibinfo
  {pages} {15} (\bibinfo {year} {2015})}\BibitemShut {NoStop}%
\bibitem [{\citenamefont {Deutsch}(1991)}]{Deustch}%
  \BibitemOpen
  \bibfield  {author} {\bibinfo {author} {\bibfnamefont {J.~M.}\ \bibnamefont
  {Deutsch}},\ }\href {\doibase 10.1103/PhysRevA.43.2046} {\bibfield  {journal}
  {\bibinfo  {journal} {Phys. Rev. A}\ }\textbf {\bibinfo {volume} {43}},\
  \bibinfo {pages} {2046} (\bibinfo {year} {1991})}\BibitemShut {NoStop}%
\bibitem [{\citenamefont {Tasaki}(1998)}]{Tasaki}%
  \BibitemOpen
  \bibfield  {author} {\bibinfo {author} {\bibfnamefont {H.}~\bibnamefont
  {Tasaki}},\ }\href {\doibase 10.1103/PhysRevLett.80.1373} {\bibfield
  {journal} {\bibinfo  {journal} {Phys. Rev. Lett.}\ }\textbf {\bibinfo
  {volume} {80}},\ \bibinfo {pages} {1373} (\bibinfo {year}
  {1998})}\BibitemShut {NoStop}%
\bibitem [{\citenamefont {Srednicki}(1994)}]{Srednicki}%
  \BibitemOpen
  \bibfield  {author} {\bibinfo {author} {\bibfnamefont {M.}~\bibnamefont
  {Srednicki}},\ }\href {\doibase 10.1103/PhysRevE.50.888} {\bibfield
  {journal} {\bibinfo  {journal} {Phys. Rev. E}\ }\textbf {\bibinfo {volume}
  {50}},\ \bibinfo {pages} {888} (\bibinfo {year} {1994})}\BibitemShut
  {NoStop}%
\bibitem [{\citenamefont {Rigol}\ \emph {et~al.}(2008)\citenamefont {Rigol},
  \citenamefont {Dunjko},\ and\ \citenamefont {Olshanii}}]{Rigol_Olshanii}%
  \BibitemOpen
  \bibfield  {author} {\bibinfo {author} {\bibfnamefont {M.}~\bibnamefont
  {Rigol}}, \bibinfo {author} {\bibfnamefont {V.}~\bibnamefont {Dunjko}}, \
  and\ \bibinfo {author} {\bibfnamefont {M.}~\bibnamefont {Olshanii}},\
  }\href@noop {} {\bibfield  {journal} {\bibinfo  {journal} {Nature}\ }\textbf
  {\bibinfo {volume} {452}},\ \bibinfo {pages} {854} (\bibinfo {year}
  {2008})}\BibitemShut {NoStop}%
\bibitem [{\citenamefont {Rigol}(2009)}]{Rigol_integrable}%
  \BibitemOpen
  \bibfield  {author} {\bibinfo {author} {\bibfnamefont {M.}~\bibnamefont
  {Rigol}},\ }\href {\doibase 10.1103/PhysRevLett.103.100403} {\bibfield
  {journal} {\bibinfo  {journal} {Phys. Rev. Lett.}\ }\textbf {\bibinfo
  {volume} {103}},\ \bibinfo {pages} {100403} (\bibinfo {year}
  {2009})}\BibitemShut {NoStop}%
\bibitem [{\citenamefont {Basko}\ \emph {et~al.}(2006)\citenamefont {Basko},
  \citenamefont {Aleiner},\ and\ \citenamefont {Altshuler}}]{BAA}%
  \BibitemOpen
  \bibfield  {author} {\bibinfo {author} {\bibfnamefont {D.}~\bibnamefont
  {Basko}}, \bibinfo {author} {\bibfnamefont {I.}~\bibnamefont {Aleiner}}, \
  and\ \bibinfo {author} {\bibfnamefont {B.}~\bibnamefont {Altshuler}},\ }\href
  {\doibase http://dx.doi.org/10.1016/j.aop.2005.11.014} {\bibfield  {journal}
  {\bibinfo  {journal} {Ann. Phys.}\ }\textbf {\bibinfo {volume} {321}},\
  \bibinfo {pages} {1126 } (\bibinfo {year} {2006})}\BibitemShut {NoStop}%
\bibitem [{\citenamefont {Vosk}\ \emph {et~al.}(2015)\citenamefont {Vosk},
  \citenamefont {Huse},\ and\ \citenamefont {Altman}}]{Altman_MBLtransition}%
  \BibitemOpen
  \bibfield  {author} {\bibinfo {author} {\bibfnamefont {R.}~\bibnamefont
  {Vosk}}, \bibinfo {author} {\bibfnamefont {D.~A.}\ \bibnamefont {Huse}}, \
  and\ \bibinfo {author} {\bibfnamefont {E.}~\bibnamefont {Altman}},\ }\href
  {\doibase 10.1103/PhysRevX.5.031032} {\bibfield  {journal} {\bibinfo
  {journal} {Phys. Rev. X}\ }\textbf {\bibinfo {volume} {5}},\ \bibinfo {pages}
  {031032} (\bibinfo {year} {2015})}\BibitemShut {NoStop}%
\bibitem [{\citenamefont {Khemani}\ \emph {et~al.}(2017)\citenamefont
  {Khemani}, \citenamefont {Lim}, \citenamefont {Sheng},\ and\ \citenamefont
  {Huse}}]{Khemani_MBLtransition}%
  \BibitemOpen
  \bibfield  {author} {\bibinfo {author} {\bibfnamefont {V.}~\bibnamefont
  {Khemani}}, \bibinfo {author} {\bibfnamefont {S.~P.}\ \bibnamefont {Lim}},
  \bibinfo {author} {\bibfnamefont {D.~N.}\ \bibnamefont {Sheng}}, \ and\
  \bibinfo {author} {\bibfnamefont {D.~A.}\ \bibnamefont {Huse}},\ }\href
  {\doibase 10.1103/PhysRevX.7.021013} {\bibfield  {journal} {\bibinfo
  {journal} {Phys. Rev. X}\ }\textbf {\bibinfo {volume} {7}},\ \bibinfo {pages}
  {021013} (\bibinfo {year} {2017})}\BibitemShut {NoStop}%
\bibitem [{\citenamefont {Kondov}\ \emph {et~al.}(2015)\citenamefont {Kondov},
  \citenamefont {McGehee}, \citenamefont {Xu},\ and\ \citenamefont
  {DeMarco}}]{MBL_Hubbard}%
  \BibitemOpen
  \bibfield  {author} {\bibinfo {author} {\bibfnamefont {S.~S.}\ \bibnamefont
  {Kondov}}, \bibinfo {author} {\bibfnamefont {W.~R.}\ \bibnamefont {McGehee}},
  \bibinfo {author} {\bibfnamefont {W.}~\bibnamefont {Xu}}, \ and\ \bibinfo
  {author} {\bibfnamefont {B.}~\bibnamefont {DeMarco}},\ }\href {\doibase
  10.1103/PhysRevLett.114.083002} {\bibfield  {journal} {\bibinfo  {journal}
  {Phys. Rev. Lett.}\ }\textbf {\bibinfo {volume} {114}},\ \bibinfo {pages}
  {083002} (\bibinfo {year} {2015})}\BibitemShut {NoStop}%
\bibitem [{\citenamefont {Schreiber}\ \emph {et~al.}(2015)\citenamefont
  {Schreiber}, \citenamefont {Hodgman}, \citenamefont {Bordia}, \citenamefont
  {L{\"u}schen}, \citenamefont {Fischer}, \citenamefont {Vosk}, \citenamefont
  {Altman}, \citenamefont {Schneider},\ and\ \citenamefont {Bloch}}]{MBL1D}%
  \BibitemOpen
  \bibfield  {author} {\bibinfo {author} {\bibfnamefont {M.}~\bibnamefont
  {Schreiber}}, \bibinfo {author} {\bibfnamefont {S.~S.}\ \bibnamefont
  {Hodgman}}, \bibinfo {author} {\bibfnamefont {P.}~\bibnamefont {Bordia}},
  \bibinfo {author} {\bibfnamefont {H.~P.}\ \bibnamefont {L{\"u}schen}},
  \bibinfo {author} {\bibfnamefont {M.~H.}\ \bibnamefont {Fischer}}, \bibinfo
  {author} {\bibfnamefont {R.}~\bibnamefont {Vosk}}, \bibinfo {author}
  {\bibfnamefont {E.}~\bibnamefont {Altman}}, \bibinfo {author} {\bibfnamefont
  {U.}~\bibnamefont {Schneider}}, \ and\ \bibinfo {author} {\bibfnamefont
  {I.}~\bibnamefont {Bloch}},\ }\href@noop {} {\bibfield  {journal} {\bibinfo
  {journal} {Science}\ }\textbf {\bibinfo {volume} {349}},\ \bibinfo {pages}
  {842} (\bibinfo {year} {2015})}\BibitemShut {NoStop}%
\bibitem [{\citenamefont {Smith}\ \emph {et~al.}(2016)\citenamefont {Smith},
  \citenamefont {Lee}, \citenamefont {Richerme}, \citenamefont {Neyenhuis},
  \citenamefont {Hess}, \citenamefont {Hauke}, \citenamefont {Heyl},
  \citenamefont {Huse},\ and\ \citenamefont {Monroe}}]{Monroe_MBL}%
  \BibitemOpen
  \bibfield  {author} {\bibinfo {author} {\bibfnamefont {J.}~\bibnamefont
  {Smith}}, \bibinfo {author} {\bibfnamefont {A.}~\bibnamefont {Lee}}, \bibinfo
  {author} {\bibfnamefont {P.}~\bibnamefont {Richerme}}, \bibinfo {author}
  {\bibfnamefont {B.}~\bibnamefont {Neyenhuis}}, \bibinfo {author}
  {\bibfnamefont {P.~W.}\ \bibnamefont {Hess}}, \bibinfo {author}
  {\bibfnamefont {P.}~\bibnamefont {Hauke}}, \bibinfo {author} {\bibfnamefont
  {M.}~\bibnamefont {Heyl}}, \bibinfo {author} {\bibfnamefont {D.~A.}\
  \bibnamefont {Huse}}, \ and\ \bibinfo {author} {\bibfnamefont
  {C.}~\bibnamefont {Monroe}},\ }\href@noop {} {\bibfield  {journal} {\bibinfo
  {journal} {Nat. Phys.}\ }\textbf {\bibinfo {volume} {12}},\ \bibinfo {pages}
  {907} (\bibinfo {year} {2016})}\BibitemShut {NoStop}%
\bibitem [{\citenamefont {Choi}\ \emph {et~al.}(2016)\citenamefont {Choi},
  \citenamefont {Hild}, \citenamefont {Zeiher}, \citenamefont {Schau{\ss}},
  \citenamefont {Rubio-Abadal}, \citenamefont {Yefsah}, \citenamefont
  {Khemani}, \citenamefont {Huse}, \citenamefont {Bloch},\ and\ \citenamefont
  {Gross}}]{MBL2D_2}%
  \BibitemOpen
  \bibfield  {author} {\bibinfo {author} {\bibfnamefont {J.-Y.}\ \bibnamefont
  {Choi}}, \bibinfo {author} {\bibfnamefont {S.}~\bibnamefont {Hild}}, \bibinfo
  {author} {\bibfnamefont {J.}~\bibnamefont {Zeiher}}, \bibinfo {author}
  {\bibfnamefont {P.}~\bibnamefont {Schau{\ss}}}, \bibinfo {author}
  {\bibfnamefont {A.}~\bibnamefont {Rubio-Abadal}}, \bibinfo {author}
  {\bibfnamefont {T.}~\bibnamefont {Yefsah}}, \bibinfo {author} {\bibfnamefont
  {V.}~\bibnamefont {Khemani}}, \bibinfo {author} {\bibfnamefont {D.~A.}\
  \bibnamefont {Huse}}, \bibinfo {author} {\bibfnamefont {I.}~\bibnamefont
  {Bloch}}, \ and\ \bibinfo {author} {\bibfnamefont {C.}~\bibnamefont
  {Gross}},\ }\href@noop {} {\bibfield  {journal} {\bibinfo  {journal}
  {Science}\ }\textbf {\bibinfo {volume} {352}},\ \bibinfo {pages} {1547}
  (\bibinfo {year} {2016})}\BibitemShut {NoStop}%
\bibitem [{\citenamefont {L\"uschen}\ \emph {et~al.}(2017)\citenamefont
  {L\"uschen}, \citenamefont {Bordia}, \citenamefont {Hodgman}, \citenamefont
  {Schreiber}, \citenamefont {Sarkar}, \citenamefont {Daley}, \citenamefont
  {Fischer}, \citenamefont {Altman}, \citenamefont {Bloch},\ and\ \citenamefont
  {Schneider}}]{MBLopensystem}%
  \BibitemOpen
  \bibfield  {author} {\bibinfo {author} {\bibfnamefont {H.~P.}\ \bibnamefont
  {L\"uschen}}, \bibinfo {author} {\bibfnamefont {P.}~\bibnamefont {Bordia}},
  \bibinfo {author} {\bibfnamefont {S.~S.}\ \bibnamefont {Hodgman}}, \bibinfo
  {author} {\bibfnamefont {M.}~\bibnamefont {Schreiber}}, \bibinfo {author}
  {\bibfnamefont {S.}~\bibnamefont {Sarkar}}, \bibinfo {author} {\bibfnamefont
  {A.~J.}\ \bibnamefont {Daley}}, \bibinfo {author} {\bibfnamefont {M.~H.}\
  \bibnamefont {Fischer}}, \bibinfo {author} {\bibfnamefont {E.}~\bibnamefont
  {Altman}}, \bibinfo {author} {\bibfnamefont {I.}~\bibnamefont {Bloch}}, \
  and\ \bibinfo {author} {\bibfnamefont {U.}~\bibnamefont {Schneider}},\ }\href
  {\doibase 10.1103/PhysRevX.7.011034} {\bibfield  {journal} {\bibinfo
  {journal} {Phys. Rev. X}\ }\textbf {\bibinfo {volume} {7}},\ \bibinfo {pages}
  {011034} (\bibinfo {year} {2017})}\BibitemShut {NoStop}%
\bibitem [{\citenamefont {Bordia}\ \emph {et~al.}(2017)\citenamefont {Bordia},
  \citenamefont {L{\"u}schen}, \citenamefont {Scherg}, \citenamefont
  {Gopalakrishnan}, \citenamefont {Knap}, \citenamefont {Schneider},\ and\
  \citenamefont {Bloch}}]{MBL2D_1}%
  \BibitemOpen
  \bibfield  {author} {\bibinfo {author} {\bibfnamefont {P.}~\bibnamefont
  {Bordia}}, \bibinfo {author} {\bibfnamefont {H.}~\bibnamefont {L{\"u}schen}},
  \bibinfo {author} {\bibfnamefont {S.}~\bibnamefont {Scherg}}, \bibinfo
  {author} {\bibfnamefont {S.}~\bibnamefont {Gopalakrishnan}}, \bibinfo
  {author} {\bibfnamefont {M.}~\bibnamefont {Knap}}, \bibinfo {author}
  {\bibfnamefont {U.}~\bibnamefont {Schneider}}, \ and\ \bibinfo {author}
  {\bibfnamefont {I.}~\bibnamefont {Bloch}},\ }\href@noop {} {\bibfield
  {journal} {\bibinfo  {journal} {Phys. Rev. X}\ }\textbf {\bibinfo {volume}
  {7}},\ \bibinfo {pages} {041047} (\bibinfo {year} {2017})}\BibitemShut
  {NoStop}%
\bibitem [{\citenamefont {Huse}\ \emph {et~al.}(2013)\citenamefont {Huse},
  \citenamefont {Nandkishore}, \citenamefont {Oganesyan}, \citenamefont {Pal},\
  and\ \citenamefont {Sondhi}}]{Loc_quantumorder}%
  \BibitemOpen
  \bibfield  {author} {\bibinfo {author} {\bibfnamefont {D.~A.}\ \bibnamefont
  {Huse}}, \bibinfo {author} {\bibfnamefont {R.}~\bibnamefont {Nandkishore}},
  \bibinfo {author} {\bibfnamefont {V.}~\bibnamefont {Oganesyan}}, \bibinfo
  {author} {\bibfnamefont {A.}~\bibnamefont {Pal}}, \ and\ \bibinfo {author}
  {\bibfnamefont {S.~L.}\ \bibnamefont {Sondhi}},\ }\href {\doibase
  10.1103/PhysRevB.88.014206} {\bibfield  {journal} {\bibinfo  {journal} {Phys.
  Rev. B}\ }\textbf {\bibinfo {volume} {88}},\ \bibinfo {pages} {014206}
  (\bibinfo {year} {2013})}\BibitemShut {NoStop}%
\bibitem [{\citenamefont {Vasseur}\ \emph {et~al.}(2015)\citenamefont
  {Vasseur}, \citenamefont {Parameswaran},\ and\ \citenamefont
  {Moore}}]{Moore}%
  \BibitemOpen
  \bibfield  {author} {\bibinfo {author} {\bibfnamefont {R.}~\bibnamefont
  {Vasseur}}, \bibinfo {author} {\bibfnamefont {S.~A.}\ \bibnamefont
  {Parameswaran}}, \ and\ \bibinfo {author} {\bibfnamefont {J.~E.}\
  \bibnamefont {Moore}},\ }\href@noop {} {\bibfield  {journal} {\bibinfo
  {journal} {Phys Rev B}\ }\textbf {\bibinfo {volume} {91}},\ \bibinfo {pages}
  {140202} (\bibinfo {year} {2015})}\BibitemShut {NoStop}%
\bibitem [{\citenamefont {Halpern}\ \emph {et~al.}(2017)\citenamefont
  {Halpern}, \citenamefont {White}, \citenamefont {Gopalakrishnan},\ and\
  \citenamefont {Refael}}]{MBLmobile}%
  \BibitemOpen
  \bibfield  {author} {\bibinfo {author} {\bibfnamefont {N.~Y.}\ \bibnamefont
  {Halpern}}, \bibinfo {author} {\bibfnamefont {C.~D.}\ \bibnamefont {White}},
  \bibinfo {author} {\bibfnamefont {S.}~\bibnamefont {Gopalakrishnan}}, \ and\
  \bibinfo {author} {\bibfnamefont {G.}~\bibnamefont {Refael}},\ }\href@noop {}
  {\bibfield  {journal} {\bibinfo  {journal} {arXiv:1707.07008}\ } (\bibinfo
  {year} {2017})}\BibitemShut {NoStop}%
\bibitem [{\citenamefont {Morrison}\ \emph {et~al.}(2008)\citenamefont
  {Morrison}, \citenamefont {Rennick}, \citenamefont {Keller},\ and\
  \citenamefont {Grant}}]{Morrison2008}%
  \BibitemOpen
  \bibfield  {author} {\bibinfo {author} {\bibfnamefont {J.~P.}\ \bibnamefont
  {Morrison}}, \bibinfo {author} {\bibfnamefont {C.~J.}\ \bibnamefont
  {Rennick}}, \bibinfo {author} {\bibfnamefont {J.~S.}\ \bibnamefont {Keller}},
  \ and\ \bibinfo {author} {\bibfnamefont {E.~R.}\ \bibnamefont {Grant}},\
  }\href@noop {} {\bibfield  {journal} {\bibinfo  {journal} {Phys. Rev. Lett.}\
  }\textbf {\bibinfo {volume} {101}},\ \bibinfo {pages} {205005} (\bibinfo
  {year} {2008})}\BibitemShut {NoStop}%
\bibitem [{\citenamefont {Haenel}\ \emph {et~al.}(2017)\citenamefont {Haenel},
  \citenamefont {Schulz-Weiling}, \citenamefont {Sous}, \citenamefont
  {Sadeghi}, \citenamefont {Aghigh}, \citenamefont {Melo}, \citenamefont
  {Keller},\ and\ \citenamefont {Grant}}]{Haenel}%
  \BibitemOpen
  \bibfield  {author} {\bibinfo {author} {\bibfnamefont {R.}~\bibnamefont
  {Haenel}}, \bibinfo {author} {\bibfnamefont {M.}~\bibnamefont
  {Schulz-Weiling}}, \bibinfo {author} {\bibfnamefont {J.}~\bibnamefont
  {Sous}}, \bibinfo {author} {\bibfnamefont {H.}~\bibnamefont {Sadeghi}},
  \bibinfo {author} {\bibfnamefont {M.}~\bibnamefont {Aghigh}}, \bibinfo
  {author} {\bibfnamefont {L.}~\bibnamefont {Melo}}, \bibinfo {author}
  {\bibfnamefont {J.}~\bibnamefont {Keller}}, \ and\ \bibinfo {author}
  {\bibfnamefont {E.}~\bibnamefont {Grant}},\ }\href@noop {} {\bibfield
  {journal} {\bibinfo  {journal} {Phys Rev A}\ }\textbf {\bibinfo {volume}
  {96}},\ \bibinfo {pages} {023613} (\bibinfo {year} {2017})}\BibitemShut
  {NoStop}%
\bibitem [{\citenamefont {Schulz-Weiling}\ and\ \citenamefont
  {Grant}(2016)}]{SchulzWeiling}%
  \BibitemOpen
  \bibfield  {author} {\bibinfo {author} {\bibfnamefont {M.}~\bibnamefont
  {Schulz-Weiling}}\ and\ \bibinfo {author} {\bibfnamefont {E.~R.}\
  \bibnamefont {Grant}},\ }\href@noop {} {\bibfield  {journal} {\bibinfo
  {journal} {J Phys B}\ }\textbf {\bibinfo {volume} {49}},\ \bibinfo {pages}
  {064009} (\bibinfo {year} {2016})}\BibitemShut {NoStop}%
\bibitem [{\citenamefont {Schulz-Weiling}\ \emph {et~al.}(2016)\citenamefont
  {Schulz-Weiling}, \citenamefont {Sadeghi}, \citenamefont {Hung},\ and\
  \citenamefont {Grant}}]{MSW_tutorial}%
  \BibitemOpen
  \bibfield  {author} {\bibinfo {author} {\bibfnamefont {M.}~\bibnamefont
  {Schulz-Weiling}}, \bibinfo {author} {\bibfnamefont {H.}~\bibnamefont
  {Sadeghi}}, \bibinfo {author} {\bibfnamefont {J.}~\bibnamefont {Hung}}, \
  and\ \bibinfo {author} {\bibfnamefont {E.~R.}\ \bibnamefont {Grant}},\
  }\href@noop {} {\bibfield  {journal} {\bibinfo  {journal} {J Phys B}\
  }\textbf {\bibinfo {volume} {49}},\ \bibinfo {pages} {193001} (\bibinfo
  {year} {2016})}\BibitemShut {NoStop}%
\bibitem [{Sup()}]{Supp}%
  \BibitemOpen
  \href@noop {} {}\BibitemShut {NoStop}%
\bibitem [{\citenamefont {Sadeghi}\ \emph {et~al.}(2014)\citenamefont
  {Sadeghi}, \citenamefont {Kruyen}, \citenamefont {Hung}, \citenamefont
  {Gurian}, \citenamefont {Morrison}, \citenamefont {Schulz-Weiling},
  \citenamefont {Saquet}, \citenamefont {Rennick},\ and\ \citenamefont
  {Grant}}]{Sadeghi:2014}%
  \BibitemOpen
  \bibfield  {author} {\bibinfo {author} {\bibfnamefont {H.}~\bibnamefont
  {Sadeghi}}, \bibinfo {author} {\bibfnamefont {A.}~\bibnamefont {Kruyen}},
  \bibinfo {author} {\bibfnamefont {J.}~\bibnamefont {Hung}}, \bibinfo {author}
  {\bibfnamefont {J.~H.}\ \bibnamefont {Gurian}}, \bibinfo {author}
  {\bibfnamefont {J.~P.}\ \bibnamefont {Morrison}}, \bibinfo {author}
  {\bibfnamefont {M.}~\bibnamefont {Schulz-Weiling}}, \bibinfo {author}
  {\bibfnamefont {N.}~\bibnamefont {Saquet}}, \bibinfo {author} {\bibfnamefont
  {C.~J.}\ \bibnamefont {Rennick}}, \ and\ \bibinfo {author} {\bibfnamefont
  {E.~R.}\ \bibnamefont {Grant}},\ }\href@noop {} {\bibfield  {journal}
  {\bibinfo  {journal} {Phys Rev Lett}\ }\textbf {\bibinfo {volume} {112}},\
  \bibinfo {pages} {075001} (\bibinfo {year} {2014})}\BibitemShut {NoStop}%
\bibitem [{\citenamefont {Walz-Flannigan}\ \emph {et~al.}(2004)\citenamefont
  {Walz-Flannigan}, \citenamefont {Guest}, \citenamefont {Choi},\ and\
  \citenamefont {Raithel}}]{WalzFlannigan}%
  \BibitemOpen
  \bibfield  {author} {\bibinfo {author} {\bibfnamefont {A.}~\bibnamefont
  {Walz-Flannigan}}, \bibinfo {author} {\bibfnamefont {J.~R.}\ \bibnamefont
  {Guest}}, \bibinfo {author} {\bibfnamefont {J.~H.}\ \bibnamefont {Choi}}, \
  and\ \bibinfo {author} {\bibfnamefont {G.}~\bibnamefont {Raithel}},\
  }\href@noop {} {\bibfield  {journal} {\bibinfo  {journal} {Phys Rev A A}\
  }\textbf {\bibinfo {volume} {69}},\ \bibinfo {pages} {063405} (\bibinfo
  {year} {2004})}\BibitemShut {NoStop}%
\bibitem [{\citenamefont {Schneider}\ \emph {et~al.}(2000)\citenamefont
  {Schneider}, \citenamefont {Rabad{\'a}n}, \citenamefont {Carata},
  \citenamefont {Andersen}, \citenamefont {Suzor-Weiner},\ and\ \citenamefont
  {Tennyson}}]{Schneider}%
  \BibitemOpen
  \bibfield  {author} {\bibinfo {author} {\bibfnamefont {I.~F.}\ \bibnamefont
  {Schneider}}, \bibinfo {author} {\bibfnamefont {I.}~\bibnamefont
  {Rabad{\'a}n}}, \bibinfo {author} {\bibfnamefont {L.}~\bibnamefont {Carata}},
  \bibinfo {author} {\bibfnamefont {L.}~\bibnamefont {Andersen}}, \bibinfo
  {author} {\bibfnamefont {A.}~\bibnamefont {Suzor-Weiner}}, \ and\ \bibinfo
  {author} {\bibfnamefont {J.}~\bibnamefont {Tennyson}},\ }\href@noop {}
  {\bibfield  {journal} {\bibinfo  {journal} {J Phys B}\ }\textbf {\bibinfo
  {volume} {33}},\ \bibinfo {pages} {4849} (\bibinfo {year}
  {2000})}\BibitemShut {NoStop}%
\bibitem [{\citenamefont {Remacle}\ and\ \citenamefont
  {Vrakking}(1998)}]{Remacle}%
  \BibitemOpen
  \bibfield  {author} {\bibinfo {author} {\bibfnamefont {F.}~\bibnamefont
  {Remacle}}\ and\ \bibinfo {author} {\bibfnamefont {M.}~\bibnamefont
  {Vrakking}},\ }\href@noop {} {\bibfield  {journal} {\bibinfo  {journal} {J
  Phys Chem A}\ }\textbf {\bibinfo {volume} {102}},\ \bibinfo {pages} {9507}
  (\bibinfo {year} {1998})}\BibitemShut {NoStop}%
\bibitem [{\citenamefont {Greene}\ and\ \citenamefont {Jungen}(1985)}]{Greene}%
  \BibitemOpen
  \bibfield  {author} {\bibinfo {author} {\bibfnamefont {C.~H.}\ \bibnamefont
  {Greene}}\ and\ \bibinfo {author} {\bibfnamefont {C.}~\bibnamefont
  {Jungen}},\ }in\ \href@noop {} {\emph {\bibinfo {booktitle} {Adv in Atomic
  and Molecular Phys Volume 21}}}\ (\bibinfo  {publisher} {Elsevier},\ \bibinfo
  {year} {1985})\ pp.\ \bibinfo {pages} {51--121}\BibitemShut {NoStop}%
\bibitem [{\citenamefont {Baranov}\ \emph {et~al.}(2012)\citenamefont
  {Baranov}, \citenamefont {Dalmonte}, \citenamefont {Pupillo},\ and\
  \citenamefont {Zoller}}]{Pupillo}%
  \BibitemOpen
  \bibfield  {author} {\bibinfo {author} {\bibfnamefont {M.}~\bibnamefont
  {Baranov}}, \bibinfo {author} {\bibfnamefont {M.}~\bibnamefont {Dalmonte}},
  \bibinfo {author} {\bibfnamefont {G.}~\bibnamefont {Pupillo}}, \ and\
  \bibinfo {author} {\bibfnamefont {P.}~\bibnamefont {Zoller}},\ }\href@noop {}
  {\bibfield  {journal} {\bibinfo  {journal} {Chemical Reviews}\ }\textbf
  {\bibinfo {volume} {112}},\ \bibinfo {pages} {5012} (\bibinfo {year}
  {2012})}\BibitemShut {NoStop}%
\bibitem [{\citenamefont {Krems}\ \emph {et~al.}(2009)\citenamefont {Krems},
  \citenamefont {Friedrich},\ and\ \citenamefont {Stwalley}}]{Krems_book1}%
  \BibitemOpen
  \bibfield  {author} {\bibinfo {author} {\bibfnamefont {R.}~\bibnamefont
  {Krems}}, \bibinfo {author} {\bibfnamefont {B.}~\bibnamefont {Friedrich}}, \
  and\ \bibinfo {author} {\bibfnamefont {W.~C.}\ \bibnamefont {Stwalley}},\
  }\href@noop {} {\emph {\bibinfo {title} {Cold molecules: theory, experiment,
  applications}}}\ (\bibinfo  {publisher} {CRC press},\ \bibinfo {year}
  {2009})\BibitemShut {NoStop}%
\bibitem [{\citenamefont {Pillet}\ and\ \citenamefont
  {Comparat}(2010)}]{Pillet}%
  \BibitemOpen
  \bibfield  {author} {\bibinfo {author} {\bibfnamefont {P.}~\bibnamefont
  {Pillet}}\ and\ \bibinfo {author} {\bibfnamefont {D.}~\bibnamefont
  {Comparat}},\ }\href@noop {} {\bibfield  {journal} {\bibinfo  {journal} {J
  Opt Soc Am B}\ }\textbf {\bibinfo {volume} {27}},\ \bibinfo {pages} {A208}
  (\bibinfo {year} {2010})}\BibitemShut {NoStop}%
\bibitem [{\citenamefont {L{\"o}w}\ \emph {et~al.}(2012)\citenamefont
  {L{\"o}w}, \citenamefont {Weimer}, \citenamefont {Nipper}, \citenamefont
  {Balewski}, \citenamefont {Butscher}, \citenamefont {B{\"u}chler},\ and\
  \citenamefont {Pfau}}]{Low}%
  \BibitemOpen
  \bibfield  {author} {\bibinfo {author} {\bibfnamefont {R.}~\bibnamefont
  {L{\"o}w}}, \bibinfo {author} {\bibfnamefont {H.}~\bibnamefont {Weimer}},
  \bibinfo {author} {\bibfnamefont {J.}~\bibnamefont {Nipper}}, \bibinfo
  {author} {\bibfnamefont {J.~B.}\ \bibnamefont {Balewski}}, \bibinfo {author}
  {\bibfnamefont {B.}~\bibnamefont {Butscher}}, \bibinfo {author}
  {\bibfnamefont {H.~P.}\ \bibnamefont {B{\"u}chler}}, \ and\ \bibinfo {author}
  {\bibfnamefont {T.}~\bibnamefont {Pfau}},\ }\href@noop {} {\bibfield
  {journal} {\bibinfo  {journal} {J Phys B}\ }\textbf {\bibinfo {volume}
  {45}},\ \bibinfo {pages} {113001} (\bibinfo {year} {2012})}\BibitemShut
  {NoStop}%
\bibitem [{\citenamefont {Firstenberg}\ \emph {et~al.}(2016)\citenamefont
  {Firstenberg}, \citenamefont {Adams},\ and\ \citenamefont
  {Hofferberth}}]{Firstenberg}%
  \BibitemOpen
  \bibfield  {author} {\bibinfo {author} {\bibfnamefont {O.}~\bibnamefont
  {Firstenberg}}, \bibinfo {author} {\bibfnamefont {C.~S.}\ \bibnamefont
  {Adams}}, \ and\ \bibinfo {author} {\bibfnamefont {S.}~\bibnamefont
  {Hofferberth}},\ }\href@noop {} {\bibfield  {journal} {\bibinfo  {journal} {J
  Phys B}\ }\textbf {\bibinfo {volume} {49}},\ \bibinfo {pages} {152003}
  (\bibinfo {year} {2016})}\BibitemShut {NoStop}%
\bibitem [{\citenamefont {M\"ulken}\ \emph {et~al.}(2007)\citenamefont
  {M\"ulken}, \citenamefont {Blumen}, \citenamefont {Amthor}, \citenamefont
  {Giese}, \citenamefont {Reetz-Lamour},\ and\ \citenamefont
  {Weidem\"uller}}]{rydbergs1}%
  \BibitemOpen
  \bibfield  {author} {\bibinfo {author} {\bibfnamefont {O.}~\bibnamefont
  {M\"ulken}}, \bibinfo {author} {\bibfnamefont {A.}~\bibnamefont {Blumen}},
  \bibinfo {author} {\bibfnamefont {T.}~\bibnamefont {Amthor}}, \bibinfo
  {author} {\bibfnamefont {C.}~\bibnamefont {Giese}}, \bibinfo {author}
  {\bibfnamefont {M.}~\bibnamefont {Reetz-Lamour}}, \ and\ \bibinfo {author}
  {\bibfnamefont {M.}~\bibnamefont {Weidem\"uller}},\ }\href {\doibase
  10.1103/PhysRevLett.99.090601} {\bibfield  {journal} {\bibinfo  {journal}
  {Phys. Rev. Lett.}\ }\textbf {\bibinfo {volume} {99}},\ \bibinfo {pages}
  {090601} (\bibinfo {year} {2007})}\BibitemShut {NoStop}%
\bibitem [{\citenamefont {G{\"u}nter}\ \emph {et~al.}(2013)\citenamefont
  {G{\"u}nter}, \citenamefont {Schempp}, \citenamefont {Robert-de
  Saint-Vincent}, \citenamefont {Gavryusev}, \citenamefont {Helmrich},
  \citenamefont {Hofmann}, \citenamefont {Whitlock},\ and\ \citenamefont
  {Weidem{\"u}ller}}]{rydbergs5}%
  \BibitemOpen
  \bibfield  {author} {\bibinfo {author} {\bibfnamefont {G.}~\bibnamefont
  {G{\"u}nter}}, \bibinfo {author} {\bibfnamefont {H.}~\bibnamefont {Schempp}},
  \bibinfo {author} {\bibfnamefont {M.}~\bibnamefont {Robert-de
  Saint-Vincent}}, \bibinfo {author} {\bibfnamefont {V.}~\bibnamefont
  {Gavryusev}}, \bibinfo {author} {\bibfnamefont {S.}~\bibnamefont {Helmrich}},
  \bibinfo {author} {\bibfnamefont {C.~S.}\ \bibnamefont {Hofmann}}, \bibinfo
  {author} {\bibfnamefont {S.}~\bibnamefont {Whitlock}}, \ and\ \bibinfo
  {author} {\bibfnamefont {M.}~\bibnamefont {Weidem{\"u}ller}},\ }\href
  {\doibase 10.1126/science.1244843} {\bibfield  {journal} {\bibinfo  {journal}
  {Science}\ }\textbf {\bibinfo {volume} {342}},\ \bibinfo {pages} {954}
  (\bibinfo {year} {2013})}\BibitemShut {NoStop}%
\bibitem [{\citenamefont {Barredo}\ \emph {et~al.}(2015)\citenamefont
  {Barredo}, \citenamefont {Labuhn}, \citenamefont {Ravets}, \citenamefont
  {Lahaye}, \citenamefont {Browaeys},\ and\ \citenamefont {Adams}}]{Barredo}%
  \BibitemOpen
  \bibfield  {author} {\bibinfo {author} {\bibfnamefont {D.}~\bibnamefont
  {Barredo}}, \bibinfo {author} {\bibfnamefont {H.}~\bibnamefont {Labuhn}},
  \bibinfo {author} {\bibfnamefont {S.}~\bibnamefont {Ravets}}, \bibinfo
  {author} {\bibfnamefont {T.}~\bibnamefont {Lahaye}}, \bibinfo {author}
  {\bibfnamefont {A.}~\bibnamefont {Browaeys}}, \ and\ \bibinfo {author}
  {\bibfnamefont {C.~S.}\ \bibnamefont {Adams}},\ }\href@noop {} {\bibfield
  {journal} {\bibinfo  {journal} {Phys. Rev. Lett.}\ }\textbf {\bibinfo
  {volume} {114}},\ \bibinfo {pages} {113002} (\bibinfo {year}
  {2015})}\BibitemShut {NoStop}%
\bibitem [{\citenamefont {Zeiher}\ \emph {et~al.}(2017)\citenamefont {Zeiher},
  \citenamefont {Choi}, \citenamefont {Rubio-Abadal}, \citenamefont {Pohl},
  \citenamefont {van Bijnen}, \citenamefont {Bloch},\ and\ \citenamefont
  {Gross}}]{rydberg_Gross}%
  \BibitemOpen
  \bibfield  {author} {\bibinfo {author} {\bibfnamefont {J.}~\bibnamefont
  {Zeiher}}, \bibinfo {author} {\bibfnamefont {J.-Y.}\ \bibnamefont {Choi}},
  \bibinfo {author} {\bibfnamefont {A.}~\bibnamefont {Rubio-Abadal}}, \bibinfo
  {author} {\bibfnamefont {T.}~\bibnamefont {Pohl}}, \bibinfo {author}
  {\bibfnamefont {R.}~\bibnamefont {van Bijnen}}, \bibinfo {author}
  {\bibfnamefont {I.}~\bibnamefont {Bloch}}, \ and\ \bibinfo {author}
  {\bibfnamefont {C.}~\bibnamefont {Gross}},\ }\href@noop {} {\bibfield
  {journal} {\bibinfo  {journal} {Phys Rev X}\ }\textbf {\bibinfo {volume}
  {7}},\ \bibinfo {pages} {041063} (\bibinfo {year} {2017})}\BibitemShut
  {NoStop}%
\bibitem [{\citenamefont {Michaelsen}\ \emph {et~al.}(2017)\citenamefont
  {Michaelsen}, \citenamefont {Bastian}, \citenamefont {Carrascosa},
  \citenamefont {Meyer}, \citenamefont {Parker},\ and\ \citenamefont
  {Wester}}]{Wester}%
  \BibitemOpen
  \bibfield  {author} {\bibinfo {author} {\bibfnamefont {T.}~\bibnamefont
  {Michaelsen}}, \bibinfo {author} {\bibfnamefont {B.}~\bibnamefont {Bastian}},
  \bibinfo {author} {\bibfnamefont {E.}~\bibnamefont {Carrascosa}}, \bibinfo
  {author} {\bibfnamefont {J.}~\bibnamefont {Meyer}}, \bibinfo {author}
  {\bibfnamefont {D.~H.}\ \bibnamefont {Parker}}, \ and\ \bibinfo {author}
  {\bibfnamefont {R.}~\bibnamefont {Wester}},\ }\href@noop {} {\bibfield
  {journal} {\bibinfo  {journal} {J Chem Phys}\ }\textbf {\bibinfo {volume}
  {147}},\ \bibinfo {pages} {013940} (\bibinfo {year} {2017})}\BibitemShut
  {NoStop}%
\bibitem [{\citenamefont {Sachdev}(2007)}]{Sachdevbook}%
  \BibitemOpen
  \bibfield  {author} {\bibinfo {author} {\bibfnamefont {S.}~\bibnamefont
  {Sachdev}},\ }\href@noop {} {\emph {\bibinfo {title} {Quantum phase
  transitions}}}\ (\bibinfo  {publisher} {Wiley Online Library},\ \bibinfo
  {year} {2007})\BibitemShut {NoStop}%
\bibitem [{\citenamefont {Zoubi}(2015)}]{Zoubi2}%
  \BibitemOpen
  \bibfield  {author} {\bibinfo {author} {\bibfnamefont {H.}~\bibnamefont
  {Zoubi}},\ }\href@noop {} {\bibfield  {journal} {\bibinfo  {journal} {J.
  Phys. B}\ }\textbf {\bibinfo {volume} {48}},\ \bibinfo {pages} {185002}
  (\bibinfo {year} {2015})}\BibitemShut {NoStop}%
\bibitem [{\citenamefont {Samboy}(2017)}]{Samboy}%
  \BibitemOpen
  \bibfield  {author} {\bibinfo {author} {\bibfnamefont {N.}~\bibnamefont
  {Samboy}},\ }\href@noop {} {\bibfield  {journal} {\bibinfo  {journal} {Phys.
  Rev. A}\ }\textbf {\bibinfo {volume} {95}},\ \bibinfo {pages} {032702}
  (\bibinfo {year} {2017})}\BibitemShut {NoStop}%
\bibitem [{\citenamefont {Burin}(2015)}]{Burin1}%
  \BibitemOpen
  \bibfield  {author} {\bibinfo {author} {\bibfnamefont {A.~L.}\ \bibnamefont
  {Burin}},\ }\href {\doibase 10.1103/PhysRevB.92.104428} {\bibfield  {journal}
  {\bibinfo  {journal} {Phys. Rev. B}\ }\textbf {\bibinfo {volume} {92}},\
  \bibinfo {pages} {104428} (\bibinfo {year} {2015})}\BibitemShut {NoStop}%
\bibitem [{\citenamefont {Anderson}(1958)}]{Anderson}%
  \BibitemOpen
  \bibfield  {author} {\bibinfo {author} {\bibfnamefont {P.~W.}\ \bibnamefont
  {Anderson}},\ }\href {\doibase 10.1103/PhysRev.109.1492} {\bibfield
  {journal} {\bibinfo  {journal} {Phys. Rev.}\ }\textbf {\bibinfo {volume}
  {109}},\ \bibinfo {pages} {1492} (\bibinfo {year} {1958})}\BibitemShut
  {NoStop}%
\bibitem [{\citenamefont {Levitov}(1999)}]{Levitov}%
  \BibitemOpen
  \bibfield  {author} {\bibinfo {author} {\bibfnamefont {L.}~\bibnamefont
  {Levitov}},\ }\href {\doibase
  10.1002/(SICI)1521-3889(199911)8:7/9<697::AID-ANDP697>3.0.CO;2-W} {\bibfield
  {journal} {\bibinfo  {journal} {Ann. Phys.}\ }\textbf {\bibinfo {volume}
  {8}},\ \bibinfo {pages} {697} (\bibinfo {year} {1999})}\BibitemShut {NoStop}%
\bibitem [{\citenamefont {Deng}\ \emph {et~al.}(2016)\citenamefont {Deng},
  \citenamefont {Altshuler}, \citenamefont {Shlyapnikov},\ and\ \citenamefont
  {Santos}}]{Deng_nonergodic}%
  \BibitemOpen
  \bibfield  {author} {\bibinfo {author} {\bibfnamefont {X.}~\bibnamefont
  {Deng}}, \bibinfo {author} {\bibfnamefont {B.~L.}\ \bibnamefont {Altshuler}},
  \bibinfo {author} {\bibfnamefont {G.~V.}\ \bibnamefont {Shlyapnikov}}, \ and\
  \bibinfo {author} {\bibfnamefont {L.}~\bibnamefont {Santos}},\ }\href
  {\doibase 10.1103/PhysRevLett.117.020401} {\bibfield  {journal} {\bibinfo
  {journal} {Phys. Rev. Lett.}\ }\textbf {\bibinfo {volume} {117}},\ \bibinfo
  {pages} {020401} (\bibinfo {year} {2016})}\BibitemShut {NoStop}%
\bibitem [{\citenamefont {Deng}\ \emph {et~al.}(2017)\citenamefont {Deng},
  \citenamefont {Kravtsov}, \citenamefont {Shlyapnikov},\ and\ \citenamefont
  {Santos}}]{Deng_algebraic}%
  \BibitemOpen
  \bibfield  {author} {\bibinfo {author} {\bibfnamefont {X.}~\bibnamefont
  {Deng}}, \bibinfo {author} {\bibfnamefont {V.}~\bibnamefont {Kravtsov}},
  \bibinfo {author} {\bibfnamefont {G.}~\bibnamefont {Shlyapnikov}}, \ and\
  \bibinfo {author} {\bibfnamefont {L.}~\bibnamefont {Santos}},\ }\href@noop {}
  {\bibfield  {journal} {\bibinfo  {journal} {arXiv:1706.04088}\ } (\bibinfo
  {year} {2017})}\BibitemShut {NoStop}%
\bibitem [{\citenamefont {Burin}(2006)}]{Burin2}%
  \BibitemOpen
  \bibfield  {author} {\bibinfo {author} {\bibfnamefont {A.~L.}\ \bibnamefont
  {Burin}},\ }\href@noop {} {\bibfield  {journal} {\bibinfo  {journal}
  {arXiv:cond-mat/0611387}\ } (\bibinfo {year} {2006})}\BibitemShut {NoStop}%
\bibitem [{\citenamefont {Yao}\ \emph {et~al.}(2014)\citenamefont {Yao},
  \citenamefont {Laumann}, \citenamefont {Gopalakrishnan}, \citenamefont
  {Knap}, \citenamefont {M\"uller}, \citenamefont {Demler},\ and\ \citenamefont
  {Lukin}}]{Yao}%
  \BibitemOpen
  \bibfield  {author} {\bibinfo {author} {\bibfnamefont {N.~Y.}\ \bibnamefont
  {Yao}}, \bibinfo {author} {\bibfnamefont {C.~R.}\ \bibnamefont {Laumann}},
  \bibinfo {author} {\bibfnamefont {S.}~\bibnamefont {Gopalakrishnan}},
  \bibinfo {author} {\bibfnamefont {M.}~\bibnamefont {Knap}}, \bibinfo {author}
  {\bibfnamefont {M.}~\bibnamefont {M\"uller}}, \bibinfo {author}
  {\bibfnamefont {E.~A.}\ \bibnamefont {Demler}}, \ and\ \bibinfo {author}
  {\bibfnamefont {M.~D.}\ \bibnamefont {Lukin}},\ }\href {\doibase
  10.1103/PhysRevLett.113.243002} {\bibfield  {journal} {\bibinfo  {journal}
  {Phys. Rev. Lett.}\ }\textbf {\bibinfo {volume} {113}},\ \bibinfo {pages}
  {243002} (\bibinfo {year} {2014})}\BibitemShut {NoStop}%
\bibitem [{\citenamefont {Nandkishore}\ and\ \citenamefont
  {Sondhi}(2017)}]{Sondhi}%
  \BibitemOpen
  \bibfield  {author} {\bibinfo {author} {\bibfnamefont {R.~M.}\ \bibnamefont
  {Nandkishore}}\ and\ \bibinfo {author} {\bibfnamefont {S.}~\bibnamefont
  {Sondhi}},\ }\href@noop {} {\bibfield  {journal} {\bibinfo  {journal} {Phys
  Rev X}\ }\textbf {\bibinfo {volume} {7}},\ \bibinfo {pages} {041021}
  (\bibinfo {year} {2017})}\BibitemShut {NoStop}%
\bibitem [{\citenamefont {Kucsko}\ \emph {et~al.}(2016)\citenamefont {Kucsko},
  \citenamefont {Choi}, \citenamefont {Choi}, \citenamefont {Maurer},
  \citenamefont {Sumiya}, \citenamefont {Onoda}, \citenamefont {Isoya},
  \citenamefont {Jelezko}, \citenamefont {Demler}, \citenamefont {Yao} \emph
  {et~al.}}]{Lukin_critical}%
  \BibitemOpen
  \bibfield  {author} {\bibinfo {author} {\bibfnamefont {G.}~\bibnamefont
  {Kucsko}}, \bibinfo {author} {\bibfnamefont {S.}~\bibnamefont {Choi}},
  \bibinfo {author} {\bibfnamefont {J.}~\bibnamefont {Choi}}, \bibinfo {author}
  {\bibfnamefont {P.~C.}\ \bibnamefont {Maurer}}, \bibinfo {author}
  {\bibfnamefont {H.}~\bibnamefont {Sumiya}}, \bibinfo {author} {\bibfnamefont
  {S.}~\bibnamefont {Onoda}}, \bibinfo {author} {\bibfnamefont
  {J.}~\bibnamefont {Isoya}}, \bibinfo {author} {\bibfnamefont
  {F.}~\bibnamefont {Jelezko}}, \bibinfo {author} {\bibfnamefont
  {E.}~\bibnamefont {Demler}}, \bibinfo {author} {\bibfnamefont {N.~Y.}\
  \bibnamefont {Yao}},  \emph {et~al.},\ }\href@noop {} {\bibfield  {journal}
  {\bibinfo  {journal} {arXiv:1609.08216}\ } (\bibinfo {year}
  {2016})}\BibitemShut {NoStop}%
\bibitem [{\citenamefont {Chandran}\ \emph {et~al.}(2016)\citenamefont
  {Chandran}, \citenamefont {Pal}, \citenamefont {Laumann},\ and\ \citenamefont
  {Scardicchio}}]{Scardicchio_allD}%
  \BibitemOpen
  \bibfield  {author} {\bibinfo {author} {\bibfnamefont {A.}~\bibnamefont
  {Chandran}}, \bibinfo {author} {\bibfnamefont {A.}~\bibnamefont {Pal}},
  \bibinfo {author} {\bibfnamefont {C.~R.}\ \bibnamefont {Laumann}}, \ and\
  \bibinfo {author} {\bibfnamefont {A.}~\bibnamefont {Scardicchio}},\ }\href
  {\doibase 10.1103/PhysRevB.94.144203} {\bibfield  {journal} {\bibinfo
  {journal} {Phys. Rev. B}\ }\textbf {\bibinfo {volume} {94}},\ \bibinfo
  {pages} {144203} (\bibinfo {year} {2016})}\BibitemShut {NoStop}%
\bibitem [{\citenamefont {Inglis}\ and\ \citenamefont
  {Pollet}(2016)}]{Pollet_MBL}%
  \BibitemOpen
  \bibfield  {author} {\bibinfo {author} {\bibfnamefont {S.}~\bibnamefont
  {Inglis}}\ and\ \bibinfo {author} {\bibfnamefont {L.}~\bibnamefont
  {Pollet}},\ }\href {\doibase 10.1103/PhysRevLett.117.120402} {\bibfield
  {journal} {\bibinfo  {journal} {Phys. Rev. Lett.}\ }\textbf {\bibinfo
  {volume} {117}},\ \bibinfo {pages} {120402} (\bibinfo {year}
  {2016})}\BibitemShut {NoStop}%
\bibitem [{\citenamefont {De~Roeck}\ and\ \citenamefont
  {Huveneers}(2017)}]{Roeck_griffith}%
  \BibitemOpen
  \bibfield  {author} {\bibinfo {author} {\bibfnamefont {W.}~\bibnamefont
  {De~Roeck}}\ and\ \bibinfo {author} {\bibfnamefont {F.}~\bibnamefont
  {Huveneers}},\ }\href {\doibase 10.1103/PhysRevB.95.155129} {\bibfield
  {journal} {\bibinfo  {journal} {Phys. Rev. B}\ }\textbf {\bibinfo {volume}
  {95}},\ \bibinfo {pages} {155129} (\bibinfo {year} {2017})}\BibitemShut
  {NoStop}%
\bibitem [{\citenamefont {Varma}\ \emph {et~al.}(2017)\citenamefont {Varma},
  \citenamefont {Lerose}, \citenamefont {Pietracaprina}, \citenamefont
  {Goold},\ and\ \citenamefont {Scardicchio}}]{Scardicchio_ergodicity}%
  \BibitemOpen
  \bibfield  {author} {\bibinfo {author} {\bibfnamefont {V.~K.}\ \bibnamefont
  {Varma}}, \bibinfo {author} {\bibfnamefont {A.}~\bibnamefont {Lerose}},
  \bibinfo {author} {\bibfnamefont {F.}~\bibnamefont {Pietracaprina}}, \bibinfo
  {author} {\bibfnamefont {J.}~\bibnamefont {Goold}}, \ and\ \bibinfo {author}
  {\bibfnamefont {A.}~\bibnamefont {Scardicchio}},\ }\href@noop {} {\bibfield
  {journal} {\bibinfo  {journal} {Journal of Statistical Mechanics: Theory and
  Experiment}\ }\textbf {\bibinfo {volume} {2017}},\ \bibinfo {pages} {053101}
  (\bibinfo {year} {2017})}\BibitemShut {NoStop}%
\bibitem [{\citenamefont {Gallagher}(2005)}]{Gallagher}%
  \BibitemOpen
  \bibfield  {author} {\bibinfo {author} {\bibfnamefont {T.~F.}\ \bibnamefont
  {Gallagher}},\ }\href@noop {} {\emph {\bibinfo {title} {Rydberg Atoms}}}\
  (\bibinfo  {publisher} {Cambridge University Press},\ \bibinfo {year}
  {2005})\BibitemShut {NoStop}%
\bibitem [{\citenamefont {Patel}\ \emph {et~al.}(2007)\citenamefont {Patel},
  \citenamefont {Jones},\ and\ \citenamefont {Fielding}}]{Fielding}%
  \BibitemOpen
  \bibfield  {author} {\bibinfo {author} {\bibfnamefont {R.}~\bibnamefont
  {Patel}}, \bibinfo {author} {\bibfnamefont {N.}~\bibnamefont {Jones}}, \ and\
  \bibinfo {author} {\bibfnamefont {H.}~\bibnamefont {Fielding}},\ }\href@noop
  {} {\bibfield  {journal} {\bibinfo  {journal} {Phys Rev A}\ }\textbf
  {\bibinfo {volume} {76}},\ \bibinfo {pages} {043413} (\bibinfo {year}
  {2007})}\BibitemShut {NoStop}%
\bibitem [{\citenamefont {Mansbach}\ and\ \citenamefont
  {Keck}(1969)}]{Mansbach}%
  \BibitemOpen
  \bibfield  {author} {\bibinfo {author} {\bibfnamefont {P.}~\bibnamefont
  {Mansbach}}\ and\ \bibinfo {author} {\bibfnamefont {J.}~\bibnamefont
  {Keck}},\ }\href@noop {} {\bibfield  {journal} {\bibinfo  {journal} {Phys.
  Rev.}\ }\textbf {\bibinfo {volume} {181}},\ \bibinfo {pages} {275} (\bibinfo
  {year} {1969})}\BibitemShut {NoStop}%
\bibitem [{\citenamefont {Pohl}\ \emph {et~al.}(2008)\citenamefont {Pohl},
  \citenamefont {Vrinceanu},\ and\ \citenamefont {Sadeghpour}}]{PVS}%
  \BibitemOpen
  \bibfield  {author} {\bibinfo {author} {\bibfnamefont {T.}~\bibnamefont
  {Pohl}}, \bibinfo {author} {\bibfnamefont {D.}~\bibnamefont {Vrinceanu}}, \
  and\ \bibinfo {author} {\bibfnamefont {H.~R.}\ \bibnamefont {Sadeghpour}},\
  }\href@noop {} {\bibfield  {journal} {\bibinfo  {journal} {Phys. Rev. Lett.}\
  }\textbf {\bibinfo {volume} {100}},\ \bibinfo {pages} {223201} (\bibinfo
  {year} {2008})}\BibitemShut {NoStop}%
\bibitem [{\citenamefont {Bixon}\ and\ \citenamefont {Jortner}(1996)}]{Bixon}%
  \BibitemOpen
  \bibfield  {author} {\bibinfo {author} {\bibfnamefont {M.}~\bibnamefont
  {Bixon}}\ and\ \bibinfo {author} {\bibfnamefont {J.}~\bibnamefont
  {Jortner}},\ }\href@noop {} {\bibfield  {journal} {\bibinfo  {journal}
  {Journal of Modern Optics}\ }\textbf {\bibinfo {volume} {89}},\ \bibinfo
  {pages} {373} (\bibinfo {year} {1996})}\BibitemShut {NoStop}%
\bibitem [{\citenamefont {Murgu}\ \emph {et~al.}(2001)\citenamefont {Murgu},
  \citenamefont {Martin},\ and\ \citenamefont {Gallagher}}]{GallagherNO}%
  \BibitemOpen
  \bibfield  {author} {\bibinfo {author} {\bibfnamefont {E.}~\bibnamefont
  {Murgu}}, \bibinfo {author} {\bibfnamefont {J.~D.~D.}\ \bibnamefont
  {Martin}}, \ and\ \bibinfo {author} {\bibfnamefont {T.~F.}\ \bibnamefont
  {Gallagher}},\ }\href@noop {} {\bibfield  {journal} {\bibinfo  {journal} {J.
  Chem. Phys.}\ }\textbf {\bibinfo {volume} {115}},\ \bibinfo {pages} {7032}
  (\bibinfo {year} {2001})}\BibitemShut {NoStop}%
\bibitem [{\citenamefont {Chupka}(1993)}]{Chupka:1993}%
  \BibitemOpen
  \bibfield  {author} {\bibinfo {author} {\bibfnamefont {W.~A.}\ \bibnamefont
  {Chupka}},\ }\href@noop {} {\bibfield  {journal} {\bibinfo  {journal} {J Chem
  Phys}\ }\textbf {\bibinfo {volume} {98}},\ \bibinfo {pages} {4520} (\bibinfo
  {year} {1993})}\BibitemShut {NoStop}%
\bibitem [{\citenamefont {Saquet}\ \emph {et~al.}(2011)\citenamefont {Saquet},
  \citenamefont {Morrison}, \citenamefont {Schulz-Weiling}, \citenamefont
  {Sadeghi}, \citenamefont {Yiu}, \citenamefont {Rennick},\ and\ \citenamefont
  {Grant}}]{Saquet2011}%
  \BibitemOpen
  \bibfield  {author} {\bibinfo {author} {\bibfnamefont {N.}~\bibnamefont
  {Saquet}}, \bibinfo {author} {\bibfnamefont {J.~P.}\ \bibnamefont
  {Morrison}}, \bibinfo {author} {\bibfnamefont {M.}~\bibnamefont
  {Schulz-Weiling}}, \bibinfo {author} {\bibfnamefont {H.}~\bibnamefont
  {Sadeghi}}, \bibinfo {author} {\bibfnamefont {J.}~\bibnamefont {Yiu}},
  \bibinfo {author} {\bibfnamefont {C.~J.}\ \bibnamefont {Rennick}}, \ and\
  \bibinfo {author} {\bibfnamefont {E.~R.}\ \bibnamefont {Grant}},\ }\href@noop
  {} {\bibfield  {journal} {\bibinfo  {journal} {J Phys B}\ }\textbf {\bibinfo
  {volume} {44}},\ \bibinfo {pages} {184015} (\bibinfo {year}
  {2011})}\BibitemShut {NoStop}%
\bibitem [{\citenamefont {Saquet}\ \emph {et~al.}(2012)\citenamefont {Saquet},
  \citenamefont {Morrison},\ and\ \citenamefont {Grant}}]{Saquet2012}%
  \BibitemOpen
  \bibfield  {author} {\bibinfo {author} {\bibfnamefont {N.}~\bibnamefont
  {Saquet}}, \bibinfo {author} {\bibfnamefont {J.~P.}\ \bibnamefont
  {Morrison}}, \ and\ \bibinfo {author} {\bibfnamefont {E.}~\bibnamefont
  {Grant}},\ }\href@noop {} {\bibfield  {journal} {\bibinfo  {journal} {J Phys
  B}\ }\textbf {\bibinfo {volume} {45}},\ \bibinfo {pages} {175302} (\bibinfo
  {year} {2012})}\BibitemShut {NoStop}%
\bibitem [{\citenamefont {Dorozhkina}\ and\ \citenamefont
  {Semenov}(1998)}]{Dorozhkina}%
  \BibitemOpen
  \bibfield  {author} {\bibinfo {author} {\bibfnamefont {D.~S.}\ \bibnamefont
  {Dorozhkina}}\ and\ \bibinfo {author} {\bibfnamefont {V.~E.}\ \bibnamefont
  {Semenov}},\ }\href@noop {} {\bibfield  {journal} {\bibinfo  {journal} {Exact
  solutions for matter-enhanced neutrino oscillations}\ }\textbf {\bibinfo
  {volume} {81}},\ \bibinfo {pages} {2691} (\bibinfo {year}
  {1998})}\BibitemShut {NoStop}%
\bibitem [{\citenamefont {Sadeghi}\ and\ \citenamefont
  {Grant}(2012)}]{Sadeghi:2012}%
  \BibitemOpen
  \bibfield  {author} {\bibinfo {author} {\bibfnamefont {H.}~\bibnamefont
  {Sadeghi}}\ and\ \bibinfo {author} {\bibfnamefont {E.~R.}\ \bibnamefont
  {Grant}},\ }\href@noop {} {\bibfield  {journal} {\bibinfo  {journal} {Phys
  Rev A}\ }\textbf {\bibinfo {volume} {86}},\ \bibinfo {pages} {052701}
  (\bibinfo {year} {2012})}\BibitemShut {NoStop}%
\bibitem [{\citenamefont {Agranovich}(2009)}]{agranovich}%
  \BibitemOpen
  \bibfield  {author} {\bibinfo {author} {\bibfnamefont {V.~M.}\ \bibnamefont
  {Agranovich}},\ }\href@noop {} {\emph {\bibinfo {title} {Excitations in
  organic solids}}},\ Vol.\ \bibinfo {volume} {142}\ (\bibinfo  {publisher}
  {Oxford: Oxford University Press},\ \bibinfo {year} {2009})\BibitemShut
  {NoStop}%
\bibitem [{\citenamefont {Brown}\ and\ \citenamefont
  {Carrington}(2003)}]{BrownC}%
  \BibitemOpen
  \bibfield  {author} {\bibinfo {author} {\bibfnamefont {J.~M.}\ \bibnamefont
  {Brown}}\ and\ \bibinfo {author} {\bibfnamefont {A.}~\bibnamefont
  {Carrington}},\ }\href@noop {} {\emph {\bibinfo {title} {Rotational
  spectroscopy of diatomic molecules}}}\ (\bibinfo  {publisher} {Cambridge
  University Press},\ \bibinfo {year} {2003})\BibitemShut {NoStop}%
\bibitem [{\citenamefont {Gurian}\ \emph {et~al.}(2012)\citenamefont {Gurian},
  \citenamefont {Cheinet}, \citenamefont {Huillery}, \citenamefont {Fioretti},
  \citenamefont {Zhao}, \citenamefont {Gould}, \citenamefont {Comparat},\ and\
  \citenamefont {Pillet}}]{Pillet_Cs}%
  \BibitemOpen
  \bibfield  {author} {\bibinfo {author} {\bibfnamefont {J.~H.}\ \bibnamefont
  {Gurian}}, \bibinfo {author} {\bibfnamefont {P.}~\bibnamefont {Cheinet}},
  \bibinfo {author} {\bibfnamefont {P.}~\bibnamefont {Huillery}}, \bibinfo
  {author} {\bibfnamefont {A.}~\bibnamefont {Fioretti}}, \bibinfo {author}
  {\bibfnamefont {J.}~\bibnamefont {Zhao}}, \bibinfo {author} {\bibfnamefont
  {P.~L.}\ \bibnamefont {Gould}}, \bibinfo {author} {\bibfnamefont
  {D.}~\bibnamefont {Comparat}}, \ and\ \bibinfo {author} {\bibfnamefont
  {P.}~\bibnamefont {Pillet}},\ }\href@noop {} {\bibfield  {journal} {\bibinfo
  {journal} {Phys Rev Lett}\ }\textbf {\bibinfo {volume} {108}},\ \bibinfo
  {pages} {023005} (\bibinfo {year} {2012})}\BibitemShut {NoStop}%
\bibitem [{\citenamefont {Zoubi}\ \emph {et~al.}(2014)\citenamefont {Zoubi},
  \citenamefont {Eisfeld},\ and\ \citenamefont {W\"uster}}]{Zoubi1}%
  \BibitemOpen
  \bibfield  {author} {\bibinfo {author} {\bibfnamefont {H.}~\bibnamefont
  {Zoubi}}, \bibinfo {author} {\bibfnamefont {A.}~\bibnamefont {Eisfeld}}, \
  and\ \bibinfo {author} {\bibfnamefont {S.}~\bibnamefont {W\"uster}},\ }\href
  {\doibase 10.1103/PhysRevA.89.053426} {\bibfield  {journal} {\bibinfo
  {journal} {Phys. Rev. A}\ }\textbf {\bibinfo {volume} {89}},\ \bibinfo
  {pages} {053426} (\bibinfo {year} {2014})}\BibitemShut {NoStop}%
\bibitem [{\citenamefont {Imbrie}(2016)}]{Imbrie}%
  \BibitemOpen
  \bibfield  {author} {\bibinfo {author} {\bibfnamefont {J.~Z.}\ \bibnamefont
  {Imbrie}},\ }\href@noop {} {\bibfield  {journal} {\bibinfo  {journal}
  {Physical Review Letters}\ }\textbf {\bibinfo {volume} {117}},\ \bibinfo
  {pages} {027201} (\bibinfo {year} {2016})}\BibitemShut {NoStop}%
\bibitem [{\citenamefont {Gopalakrishnan}\ \emph {et~al.}(2015)\citenamefont
  {Gopalakrishnan}, \citenamefont {M\"uller}, \citenamefont {Khemani},
  \citenamefont {Knap}, \citenamefont {Demler},\ and\ \citenamefont
  {Huse}}]{Sarang1}%
  \BibitemOpen
  \bibfield  {author} {\bibinfo {author} {\bibfnamefont {S.}~\bibnamefont
  {Gopalakrishnan}}, \bibinfo {author} {\bibfnamefont {M.}~\bibnamefont
  {M\"uller}}, \bibinfo {author} {\bibfnamefont {V.}~\bibnamefont {Khemani}},
  \bibinfo {author} {\bibfnamefont {M.}~\bibnamefont {Knap}}, \bibinfo {author}
  {\bibfnamefont {E.}~\bibnamefont {Demler}}, \ and\ \bibinfo {author}
  {\bibfnamefont {D.~A.}\ \bibnamefont {Huse}},\ }\href {\doibase
  10.1103/PhysRevB.92.104202} {\bibfield  {journal} {\bibinfo  {journal} {Phys.
  Rev. B}\ }\textbf {\bibinfo {volume} {92}},\ \bibinfo {pages} {104202}
  (\bibinfo {year} {2015})}\BibitemShut {NoStop}%
\bibitem [{\citenamefont {Gopalakrishnan}\ \emph {et~al.}(2016)\citenamefont
  {Gopalakrishnan}, \citenamefont {Agarwal}, \citenamefont {Demler},
  \citenamefont {Huse},\ and\ \citenamefont {Knap}}]{Sarang2}%
  \BibitemOpen
  \bibfield  {author} {\bibinfo {author} {\bibfnamefont {S.}~\bibnamefont
  {Gopalakrishnan}}, \bibinfo {author} {\bibfnamefont {K.}~\bibnamefont
  {Agarwal}}, \bibinfo {author} {\bibfnamefont {E.~A.}\ \bibnamefont {Demler}},
  \bibinfo {author} {\bibfnamefont {D.~A.}\ \bibnamefont {Huse}}, \ and\
  \bibinfo {author} {\bibfnamefont {M.}~\bibnamefont {Knap}},\ }\href {\doibase
  10.1103/PhysRevB.93.134206} {\bibfield  {journal} {\bibinfo  {journal} {Phys.
  Rev. B}\ }\textbf {\bibinfo {volume} {93}},\ \bibinfo {pages} {134206}
  (\bibinfo {year} {2016})}\BibitemShut {NoStop}%
\bibitem [{\citenamefont {Agarwal}\ \emph {et~al.}(2017)\citenamefont
  {Agarwal}, \citenamefont {Altman}, \citenamefont {Demler}, \citenamefont
  {Gopalakrishnan}, \citenamefont {Huse},\ and\ \citenamefont
  {Knap}}]{RareRegions_rev}%
  \BibitemOpen
  \bibfield  {author} {\bibinfo {author} {\bibfnamefont {K.}~\bibnamefont
  {Agarwal}}, \bibinfo {author} {\bibfnamefont {E.}~\bibnamefont {Altman}},
  \bibinfo {author} {\bibfnamefont {E.}~\bibnamefont {Demler}}, \bibinfo
  {author} {\bibfnamefont {S.}~\bibnamefont {Gopalakrishnan}}, \bibinfo
  {author} {\bibfnamefont {D.~A.}\ \bibnamefont {Huse}}, \ and\ \bibinfo
  {author} {\bibfnamefont {M.}~\bibnamefont {Knap}},\ }\href {\doibase
  10.1002/andp.201600326} {\bibfield  {journal} {\bibinfo  {journal} {Ann.
  Phys.}\ }\textbf {\bibinfo {volume} {529}},\ \bibinfo {pages} {1600326}
  (\bibinfo {year} {2017})},\ \bibinfo {note} {1600326}\BibitemShut {NoStop}%
\bibitem [{\citenamefont {Ponte}\ \emph {et~al.}(2017)\citenamefont {Ponte},
  \citenamefont {Laumann}, \citenamefont {Huse},\ and\ \citenamefont
  {Chandran}}]{Thermal_inclusions}%
  \BibitemOpen
  \bibfield  {author} {\bibinfo {author} {\bibfnamefont {P.}~\bibnamefont
  {Ponte}}, \bibinfo {author} {\bibfnamefont {C.~R.}\ \bibnamefont {Laumann}},
  \bibinfo {author} {\bibfnamefont {D.~A.}\ \bibnamefont {Huse}}, \ and\
  \bibinfo {author} {\bibfnamefont {A.}~\bibnamefont {Chandran}},\ }\href@noop
  {} {\bibfield  {journal} {\bibinfo  {journal} {Phil. Trans. R. Soc. A}\
  }\textbf {\bibinfo {volume} {375}},\ \bibinfo {pages} {20160428} (\bibinfo
  {year} {2017})}\BibitemShut {NoStop}%
\end{thebibliography}%


\begin{thebibliography}{33}%
\makeatletter
\providecommand \@ifxundefined [1]{%
 \@ifx{#1\undefined}
}%
\providecommand \@ifnum [1]{%
 \ifnum #1\expandafter \@firstoftwo
 \else \expandafter \@secondoftwo
 \fi
}%
\providecommand \@ifx [1]{%
 \ifx #1\expandafter \@firstoftwo
 \else \expandafter \@secondoftwo
 \fi
}%
\providecommand \natexlab [1]{#1}%
\providecommand \enquote  [1]{``#1''}%
\providecommand \bibnamefont  [1]{#1}%
\providecommand \bibfnamefont [1]{#1}%
\providecommand \citenamefont [1]{#1}%
\providecommand \href@noop [0]{\@secondoftwo}%
\providecommand \href [0]{\begingroup \@sanitize@url \@href}%
\providecommand \@href[1]{\@@startlink{#1}\@@href}%
\providecommand \@@href[1]{\endgroup#1\@@endlink}%
\providecommand \@sanitize@url [0]{\catcode `\\12\catcode `\$12\catcode
  `\&12\catcode `\#12\catcode `\^12\catcode `\_12\catcode `\%12\relax}%
\providecommand \@@startlink[1]{}%
\providecommand \@@endlink[0]{}%
\providecommand \url  [0]{\begingroup\@sanitize@url \@url }%
\providecommand \@url [1]{\endgroup\@href {#1}{\urlprefix }}%
\providecommand \urlprefix  [0]{URL }%
\providecommand \Eprint [0]{\href }%
\providecommand \doibase [0]{http://dx.doi.org/}%
\providecommand \selectlanguage [0]{\@gobble}%
\providecommand \bibinfo  [0]{\@secondoftwo}%
\providecommand \bibfield  [0]{\@secondoftwo}%
\providecommand \translation [1]{[#1]}%
\providecommand \BibitemOpen [0]{}%
\providecommand \bibitemStop [0]{}%
\providecommand \bibitemNoStop [0]{.\EOS\space}%
\providecommand \EOS [0]{\spacefactor3000\relax}%
\providecommand \BibitemShut  [1]{\csname bibitem#1\endcsname}%
\let\auto@bib@innerbib\@empty
\bibitem [{\citenamefont {Schulz-Weiling}\ \emph {et~al.}(2016)\citenamefont
  {Schulz-Weiling}, \citenamefont {Sadeghi}, \citenamefont {Hung},\ and\
  \citenamefont {Grant}}]{MSW_tutorial}%
  \BibitemOpen
  \bibfield  {author} {\bibinfo {author} {\bibfnamefont {M.}~\bibnamefont
  {Schulz-Weiling}}, \bibinfo {author} {\bibfnamefont {H.}~\bibnamefont
  {Sadeghi}}, \bibinfo {author} {\bibfnamefont {J.}~\bibnamefont {Hung}}, \
  and\ \bibinfo {author} {\bibfnamefont {E.~R.}\ \bibnamefont {Grant}},\
  }\href@noop {} {\bibfield  {journal} {\bibinfo  {journal} {J Phys B}\
  }\textbf {\bibinfo {volume} {49}},\ \bibinfo {pages} {193001} (\bibinfo
  {year} {2016})}\BibitemShut {NoStop}%
\bibitem [{\citenamefont {Gallagher}(2005)}]{Gallagher}%
  \BibitemOpen
  \bibfield  {author} {\bibinfo {author} {\bibfnamefont {T.~F.}\ \bibnamefont
  {Gallagher}},\ }\href@noop {} {\emph {\bibinfo {title} {Rydberg Atoms}}}\
  (\bibinfo  {publisher} {Cambridge University Press},\ \bibinfo {year}
  {2005})\BibitemShut {NoStop}%
\bibitem [{\citenamefont {Patel}\ \emph {et~al.}(2007)\citenamefont {Patel},
  \citenamefont {Jones},\ and\ \citenamefont {Fielding}}]{Fielding}%
  \BibitemOpen
  \bibfield  {author} {\bibinfo {author} {\bibfnamefont {R.}~\bibnamefont
  {Patel}}, \bibinfo {author} {\bibfnamefont {N.}~\bibnamefont {Jones}}, \ and\
  \bibinfo {author} {\bibfnamefont {H.}~\bibnamefont {Fielding}},\ }\href@noop
  {} {\bibfield  {journal} {\bibinfo  {journal} {Phys Rev A}\ }\textbf
  {\bibinfo {volume} {76}},\ \bibinfo {pages} {043413} (\bibinfo {year}
  {2007})}\BibitemShut {NoStop}%
\bibitem [{\citenamefont {Mansbach}\ and\ \citenamefont
  {Keck}(1969)}]{Mansbach}%
  \BibitemOpen
  \bibfield  {author} {\bibinfo {author} {\bibfnamefont {P.}~\bibnamefont
  {Mansbach}}\ and\ \bibinfo {author} {\bibfnamefont {J.}~\bibnamefont
  {Keck}},\ }\href@noop {} {\bibfield  {journal} {\bibinfo  {journal} {Phys.
  Rev.}\ }\textbf {\bibinfo {volume} {181}},\ \bibinfo {pages} {275} (\bibinfo
  {year} {1969})}\BibitemShut {NoStop}%
\bibitem [{\citenamefont {Pohl}\ \emph {et~al.}(2008)\citenamefont {Pohl},
  \citenamefont {Vrinceanu},\ and\ \citenamefont {Sadeghpour}}]{PVS}%
  \BibitemOpen
  \bibfield  {author} {\bibinfo {author} {\bibfnamefont {T.}~\bibnamefont
  {Pohl}}, \bibinfo {author} {\bibfnamefont {D.}~\bibnamefont {Vrinceanu}}, \
  and\ \bibinfo {author} {\bibfnamefont {H.~R.}\ \bibnamefont {Sadeghpour}},\
  }\href@noop {} {\bibfield  {journal} {\bibinfo  {journal} {Phys. Rev. Lett.}\
  }\textbf {\bibinfo {volume} {100}},\ \bibinfo {pages} {223201} (\bibinfo
  {year} {2008})}\BibitemShut {NoStop}%
\bibitem [{\citenamefont {Bixon}\ and\ \citenamefont {Jortner}(1996)}]{Bixon}%
  \BibitemOpen
  \bibfield  {author} {\bibinfo {author} {\bibfnamefont {M.}~\bibnamefont
  {Bixon}}\ and\ \bibinfo {author} {\bibfnamefont {J.}~\bibnamefont
  {Jortner}},\ }\href@noop {} {\bibfield  {journal} {\bibinfo  {journal}
  {Journal of Modern Optics}\ }\textbf {\bibinfo {volume} {89}},\ \bibinfo
  {pages} {373} (\bibinfo {year} {1996})}\BibitemShut {NoStop}%
\bibitem [{\citenamefont {Murgu}\ \emph {et~al.}(2001)\citenamefont {Murgu},
  \citenamefont {Martin},\ and\ \citenamefont {Gallagher}}]{GallagherNO}%
  \BibitemOpen
  \bibfield  {author} {\bibinfo {author} {\bibfnamefont {E.}~\bibnamefont
  {Murgu}}, \bibinfo {author} {\bibfnamefont {J.~D.~D.}\ \bibnamefont
  {Martin}}, \ and\ \bibinfo {author} {\bibfnamefont {T.~F.}\ \bibnamefont
  {Gallagher}},\ }\href@noop {} {\bibfield  {journal} {\bibinfo  {journal} {J.
  Chem. Phys.}\ }\textbf {\bibinfo {volume} {115}},\ \bibinfo {pages} {7032}
  (\bibinfo {year} {2001})}\BibitemShut {NoStop}%
\bibitem [{\citenamefont {Remacle}\ and\ \citenamefont
  {Vrakking}(1998)}]{Remacle}%
  \BibitemOpen
  \bibfield  {author} {\bibinfo {author} {\bibfnamefont {F.}~\bibnamefont
  {Remacle}}\ and\ \bibinfo {author} {\bibfnamefont {M.}~\bibnamefont
  {Vrakking}},\ }\href@noop {} {\bibfield  {journal} {\bibinfo  {journal} {J
  Phys Chem A}\ }\textbf {\bibinfo {volume} {102}},\ \bibinfo {pages} {9507}
  (\bibinfo {year} {1998})}\BibitemShut {NoStop}%
\bibitem [{\citenamefont {Chupka}(1993)}]{Chupka:1993}%
  \BibitemOpen
  \bibfield  {author} {\bibinfo {author} {\bibfnamefont {W.~A.}\ \bibnamefont
  {Chupka}},\ }\href@noop {} {\bibfield  {journal} {\bibinfo  {journal} {J Chem
  Phys}\ }\textbf {\bibinfo {volume} {98}},\ \bibinfo {pages} {4520} (\bibinfo
  {year} {1993})}\BibitemShut {NoStop}%
\bibitem [{\citenamefont {Schneider}\ \emph {et~al.}(2000)\citenamefont
  {Schneider}, \citenamefont {Rabad{\'a}n}, \citenamefont {Carata},
  \citenamefont {Andersen}, \citenamefont {Suzor-Weiner},\ and\ \citenamefont
  {Tennyson}}]{Schneider}%
  \BibitemOpen
  \bibfield  {author} {\bibinfo {author} {\bibfnamefont {I.~F.}\ \bibnamefont
  {Schneider}}, \bibinfo {author} {\bibfnamefont {I.}~\bibnamefont
  {Rabad{\'a}n}}, \bibinfo {author} {\bibfnamefont {L.}~\bibnamefont {Carata}},
  \bibinfo {author} {\bibfnamefont {L.}~\bibnamefont {Andersen}}, \bibinfo
  {author} {\bibfnamefont {A.}~\bibnamefont {Suzor-Weiner}}, \ and\ \bibinfo
  {author} {\bibfnamefont {J.}~\bibnamefont {Tennyson}},\ }\href@noop {}
  {\bibfield  {journal} {\bibinfo  {journal} {J Phys B}\ }\textbf {\bibinfo
  {volume} {33}},\ \bibinfo {pages} {4849} (\bibinfo {year}
  {2000})}\BibitemShut {NoStop}%
\bibitem [{\citenamefont {Saquet}\ \emph {et~al.}(2011)\citenamefont {Saquet},
  \citenamefont {Morrison}, \citenamefont {Schulz-Weiling}, \citenamefont
  {Sadeghi}, \citenamefont {Yiu}, \citenamefont {Rennick},\ and\ \citenamefont
  {Grant}}]{Saquet2011}%
  \BibitemOpen
  \bibfield  {author} {\bibinfo {author} {\bibfnamefont {N.}~\bibnamefont
  {Saquet}}, \bibinfo {author} {\bibfnamefont {J.~P.}\ \bibnamefont
  {Morrison}}, \bibinfo {author} {\bibfnamefont {M.}~\bibnamefont
  {Schulz-Weiling}}, \bibinfo {author} {\bibfnamefont {H.}~\bibnamefont
  {Sadeghi}}, \bibinfo {author} {\bibfnamefont {J.}~\bibnamefont {Yiu}},
  \bibinfo {author} {\bibfnamefont {C.~J.}\ \bibnamefont {Rennick}}, \ and\
  \bibinfo {author} {\bibfnamefont {E.~R.}\ \bibnamefont {Grant}},\ }\href@noop
  {} {\bibfield  {journal} {\bibinfo  {journal} {J Phys B}\ }\textbf {\bibinfo
  {volume} {44}},\ \bibinfo {pages} {184015} (\bibinfo {year}
  {2011})}\BibitemShut {NoStop}%
\bibitem [{\citenamefont {Saquet}\ \emph {et~al.}(2012)\citenamefont {Saquet},
  \citenamefont {Morrison},\ and\ \citenamefont {Grant}}]{Saquet2012}%
  \BibitemOpen
  \bibfield  {author} {\bibinfo {author} {\bibfnamefont {N.}~\bibnamefont
  {Saquet}}, \bibinfo {author} {\bibfnamefont {J.~P.}\ \bibnamefont
  {Morrison}}, \ and\ \bibinfo {author} {\bibfnamefont {E.}~\bibnamefont
  {Grant}},\ }\href@noop {} {\bibfield  {journal} {\bibinfo  {journal} {J Phys
  B}\ }\textbf {\bibinfo {volume} {45}},\ \bibinfo {pages} {175302} (\bibinfo
  {year} {2012})}\BibitemShut {NoStop}%
\bibitem [{\citenamefont {Haenel}\ \emph {et~al.}(2017)\citenamefont {Haenel},
  \citenamefont {Schulz-Weiling}, \citenamefont {Sous}, \citenamefont
  {Sadeghi}, \citenamefont {Aghigh}, \citenamefont {Melo}, \citenamefont
  {Keller},\ and\ \citenamefont {Grant}}]{Haenel}%
  \BibitemOpen
  \bibfield  {author} {\bibinfo {author} {\bibfnamefont {R.}~\bibnamefont
  {Haenel}}, \bibinfo {author} {\bibfnamefont {M.}~\bibnamefont
  {Schulz-Weiling}}, \bibinfo {author} {\bibfnamefont {J.}~\bibnamefont
  {Sous}}, \bibinfo {author} {\bibfnamefont {H.}~\bibnamefont {Sadeghi}},
  \bibinfo {author} {\bibfnamefont {M.}~\bibnamefont {Aghigh}}, \bibinfo
  {author} {\bibfnamefont {L.}~\bibnamefont {Melo}}, \bibinfo {author}
  {\bibfnamefont {J.}~\bibnamefont {Keller}}, \ and\ \bibinfo {author}
  {\bibfnamefont {E.}~\bibnamefont {Grant}},\ }\href@noop {} {\bibfield
  {journal} {\bibinfo  {journal} {Phys Rev A}\ }\textbf {\bibinfo {volume}
  {96}},\ \bibinfo {pages} {023613} (\bibinfo {year} {2017})}\BibitemShut
  {NoStop}%
\bibitem [{\citenamefont {Dorozhkina}\ and\ \citenamefont
  {Semenov}(1998)}]{Dorozhkina}%
  \BibitemOpen
  \bibfield  {author} {\bibinfo {author} {\bibfnamefont {D.~S.}\ \bibnamefont
  {Dorozhkina}}\ and\ \bibinfo {author} {\bibfnamefont {V.~E.}\ \bibnamefont
  {Semenov}},\ }\href@noop {} {\bibfield  {journal} {\bibinfo  {journal} {Exact
  solutions for matter-enhanced neutrino oscillations}\ }\textbf {\bibinfo
  {volume} {81}},\ \bibinfo {pages} {2691} (\bibinfo {year}
  {1998})}\BibitemShut {NoStop}%
\bibitem [{\citenamefont {Sadeghi}\ and\ \citenamefont
  {Grant}(2012)}]{Sadeghi:2012}%
  \BibitemOpen
  \bibfield  {author} {\bibinfo {author} {\bibfnamefont {H.}~\bibnamefont
  {Sadeghi}}\ and\ \bibinfo {author} {\bibfnamefont {E.~R.}\ \bibnamefont
  {Grant}},\ }\href@noop {} {\bibfield  {journal} {\bibinfo  {journal} {Phys
  Rev A}\ }\textbf {\bibinfo {volume} {86}},\ \bibinfo {pages} {052701}
  (\bibinfo {year} {2012})}\BibitemShut {NoStop}%
\bibitem [{\citenamefont {Sachdev}(2007)}]{Sachdevbook}%
  \BibitemOpen
  \bibfield  {author} {\bibinfo {author} {\bibfnamefont {S.}~\bibnamefont
  {Sachdev}},\ }\href@noop {} {\emph {\bibinfo {title} {Quantum phase
  transitions}}}\ (\bibinfo  {publisher} {Wiley Online Library},\ \bibinfo
  {year} {2007})\BibitemShut {NoStop}%
\bibitem [{\citenamefont {Agranovich}(2009)}]{agranovich}%
  \BibitemOpen
  \bibfield  {author} {\bibinfo {author} {\bibfnamefont {V.~M.}\ \bibnamefont
  {Agranovich}},\ }\href@noop {} {\emph {\bibinfo {title} {Excitations in
  organic solids}}},\ Vol.\ \bibinfo {volume} {142}\ (\bibinfo  {publisher}
  {Oxford: Oxford University Press},\ \bibinfo {year} {2009})\BibitemShut
  {NoStop}%
\bibitem [{\citenamefont {Brown}\ and\ \citenamefont
  {Carrington}(2003)}]{BrownC}%
  \BibitemOpen
  \bibfield  {author} {\bibinfo {author} {\bibfnamefont {J.~M.}\ \bibnamefont
  {Brown}}\ and\ \bibinfo {author} {\bibfnamefont {A.}~\bibnamefont
  {Carrington}},\ }\href@noop {} {\emph {\bibinfo {title} {Rotational
  spectroscopy of diatomic molecules}}}\ (\bibinfo  {publisher} {Cambridge
  University Press},\ \bibinfo {year} {2003})\BibitemShut {NoStop}%
\bibitem [{\citenamefont {Gurian}\ \emph {et~al.}(2012)\citenamefont {Gurian},
  \citenamefont {Cheinet}, \citenamefont {Huillery}, \citenamefont {Fioretti},
  \citenamefont {Zhao}, \citenamefont {Gould}, \citenamefont {Comparat},\ and\
  \citenamefont {Pillet}}]{Pillet_Cs}%
  \BibitemOpen
  \bibfield  {author} {\bibinfo {author} {\bibfnamefont {J.~H.}\ \bibnamefont
  {Gurian}}, \bibinfo {author} {\bibfnamefont {P.}~\bibnamefont {Cheinet}},
  \bibinfo {author} {\bibfnamefont {P.}~\bibnamefont {Huillery}}, \bibinfo
  {author} {\bibfnamefont {A.}~\bibnamefont {Fioretti}}, \bibinfo {author}
  {\bibfnamefont {J.}~\bibnamefont {Zhao}}, \bibinfo {author} {\bibfnamefont
  {P.~L.}\ \bibnamefont {Gould}}, \bibinfo {author} {\bibfnamefont
  {D.}~\bibnamefont {Comparat}}, \ and\ \bibinfo {author} {\bibfnamefont
  {P.}~\bibnamefont {Pillet}},\ }\href@noop {} {\bibfield  {journal} {\bibinfo
  {journal} {Phys Rev Lett}\ }\textbf {\bibinfo {volume} {108}},\ \bibinfo
  {pages} {023005} (\bibinfo {year} {2012})}\BibitemShut {NoStop}%
\bibitem [{\citenamefont {Burin}(2015)}]{Burin1}%
  \BibitemOpen
  \bibfield  {author} {\bibinfo {author} {\bibfnamefont {A.~L.}\ \bibnamefont
  {Burin}},\ }\href {\doibase 10.1103/PhysRevB.92.104428} {\bibfield  {journal}
  {\bibinfo  {journal} {Phys. Rev. B}\ }\textbf {\bibinfo {volume} {92}},\
  \bibinfo {pages} {104428} (\bibinfo {year} {2015})}\BibitemShut {NoStop}%
\bibitem [{\citenamefont {Zoubi}\ \emph {et~al.}(2014)\citenamefont {Zoubi},
  \citenamefont {Eisfeld},\ and\ \citenamefont {W\"uster}}]{Zoubi1}%
  \BibitemOpen
  \bibfield  {author} {\bibinfo {author} {\bibfnamefont {H.}~\bibnamefont
  {Zoubi}}, \bibinfo {author} {\bibfnamefont {A.}~\bibnamefont {Eisfeld}}, \
  and\ \bibinfo {author} {\bibfnamefont {S.}~\bibnamefont {W\"uster}},\ }\href
  {\doibase 10.1103/PhysRevA.89.053426} {\bibfield  {journal} {\bibinfo
  {journal} {Phys. Rev. A}\ }\textbf {\bibinfo {volume} {89}},\ \bibinfo
  {pages} {053426} (\bibinfo {year} {2014})}\BibitemShut {NoStop}%
\bibitem [{\citenamefont {Zoubi}(2015)}]{Zoubi2}%
  \BibitemOpen
  \bibfield  {author} {\bibinfo {author} {\bibfnamefont {H.}~\bibnamefont
  {Zoubi}},\ }\href@noop {} {\bibfield  {journal} {\bibinfo  {journal} {J.
  Phys. B}\ }\textbf {\bibinfo {volume} {48}},\ \bibinfo {pages} {185002}
  (\bibinfo {year} {2015})}\BibitemShut {NoStop}%
\bibitem [{\citenamefont {Imbrie}(2016)}]{Imbrie}%
  \BibitemOpen
  \bibfield  {author} {\bibinfo {author} {\bibfnamefont {J.~Z.}\ \bibnamefont
  {Imbrie}},\ }\href@noop {} {\bibfield  {journal} {\bibinfo  {journal}
  {Physical Review Letters}\ }\textbf {\bibinfo {volume} {117}},\ \bibinfo
  {pages} {027201} (\bibinfo {year} {2016})}\BibitemShut {NoStop}%
\bibitem [{\citenamefont {Anderson}(1958)}]{Anderson}%
  \BibitemOpen
  \bibfield  {author} {\bibinfo {author} {\bibfnamefont {P.~W.}\ \bibnamefont
  {Anderson}},\ }\href {\doibase 10.1103/PhysRev.109.1492} {\bibfield
  {journal} {\bibinfo  {journal} {Phys. Rev.}\ }\textbf {\bibinfo {volume}
  {109}},\ \bibinfo {pages} {1492} (\bibinfo {year} {1958})}\BibitemShut
  {NoStop}%
\bibitem [{\citenamefont {Levitov}(1999)}]{Levitov}%
  \BibitemOpen
  \bibfield  {author} {\bibinfo {author} {\bibfnamefont {L.}~\bibnamefont
  {Levitov}},\ }\href {\doibase
  10.1002/(SICI)1521-3889(199911)8:7/9<697::AID-ANDP697>3.0.CO;2-W} {\bibfield
  {journal} {\bibinfo  {journal} {Ann. Phys.}\ }\textbf {\bibinfo {volume}
  {8}},\ \bibinfo {pages} {697} (\bibinfo {year} {1999})}\BibitemShut {NoStop}%
\bibitem [{\citenamefont {Burin}(2006)}]{Burin2}%
  \BibitemOpen
  \bibfield  {author} {\bibinfo {author} {\bibfnamefont {A.~L.}\ \bibnamefont
  {Burin}},\ }\href@noop {} {\bibfield  {journal} {\bibinfo  {journal} {arXiv
  preprint cond-mat/0611387}\ } (\bibinfo {year} {2006})}\BibitemShut {NoStop}%
\bibitem [{\citenamefont {Nandkishore}\ and\ \citenamefont
  {Sondhi}(2017)}]{Sondhi}%
  \BibitemOpen
  \bibfield  {author} {\bibinfo {author} {\bibfnamefont {R.~M.}\ \bibnamefont
  {Nandkishore}}\ and\ \bibinfo {author} {\bibfnamefont {S.}~\bibnamefont
  {Sondhi}},\ }\href@noop {} {\bibfield  {journal} {\bibinfo  {journal} {Phys
  Rev X}\ }\textbf {\bibinfo {volume} {7}},\ \bibinfo {pages} {041021}
  (\bibinfo {year} {2017})}\BibitemShut {NoStop}%
\bibitem [{\citenamefont {Gopalakrishnan}\ \emph {et~al.}(2015)\citenamefont
  {Gopalakrishnan}, \citenamefont {M\"uller}, \citenamefont {Khemani},
  \citenamefont {Knap}, \citenamefont {Demler},\ and\ \citenamefont
  {Huse}}]{Sarang1}%
  \BibitemOpen
  \bibfield  {author} {\bibinfo {author} {\bibfnamefont {S.}~\bibnamefont
  {Gopalakrishnan}}, \bibinfo {author} {\bibfnamefont {M.}~\bibnamefont
  {M\"uller}}, \bibinfo {author} {\bibfnamefont {V.}~\bibnamefont {Khemani}},
  \bibinfo {author} {\bibfnamefont {M.}~\bibnamefont {Knap}}, \bibinfo {author}
  {\bibfnamefont {E.}~\bibnamefont {Demler}}, \ and\ \bibinfo {author}
  {\bibfnamefont {D.~A.}\ \bibnamefont {Huse}},\ }\href {\doibase
  10.1103/PhysRevB.92.104202} {\bibfield  {journal} {\bibinfo  {journal} {Phys.
  Rev. B}\ }\textbf {\bibinfo {volume} {92}},\ \bibinfo {pages} {104202}
  (\bibinfo {year} {2015})}\BibitemShut {NoStop}%
\bibitem [{\citenamefont {Gopalakrishnan}\ \emph {et~al.}(2016)\citenamefont
  {Gopalakrishnan}, \citenamefont {Agarwal}, \citenamefont {Demler},
  \citenamefont {Huse},\ and\ \citenamefont {Knap}}]{Sarang2}%
  \BibitemOpen
  \bibfield  {author} {\bibinfo {author} {\bibfnamefont {S.}~\bibnamefont
  {Gopalakrishnan}}, \bibinfo {author} {\bibfnamefont {K.}~\bibnamefont
  {Agarwal}}, \bibinfo {author} {\bibfnamefont {E.~A.}\ \bibnamefont {Demler}},
  \bibinfo {author} {\bibfnamefont {D.~A.}\ \bibnamefont {Huse}}, \ and\
  \bibinfo {author} {\bibfnamefont {M.}~\bibnamefont {Knap}},\ }\href {\doibase
  10.1103/PhysRevB.93.134206} {\bibfield  {journal} {\bibinfo  {journal} {Phys.
  Rev. B}\ }\textbf {\bibinfo {volume} {93}},\ \bibinfo {pages} {134206}
  (\bibinfo {year} {2016})}\BibitemShut {NoStop}%
\bibitem [{\citenamefont {De~Roeck}\ and\ \citenamefont
  {Huveneers}(2017)}]{Roeck_griffith}%
  \BibitemOpen
  \bibfield  {author} {\bibinfo {author} {\bibfnamefont {W.}~\bibnamefont
  {De~Roeck}}\ and\ \bibinfo {author} {\bibfnamefont {F.}~\bibnamefont
  {Huveneers}},\ }\href {\doibase 10.1103/PhysRevB.95.155129} {\bibfield
  {journal} {\bibinfo  {journal} {Phys. Rev. B}\ }\textbf {\bibinfo {volume}
  {95}},\ \bibinfo {pages} {155129} (\bibinfo {year} {2017})}\BibitemShut
  {NoStop}%
\bibitem [{\citenamefont {Agarwal}\ \emph {et~al.}(2017)\citenamefont
  {Agarwal}, \citenamefont {Altman}, \citenamefont {Demler}, \citenamefont
  {Gopalakrishnan}, \citenamefont {Huse},\ and\ \citenamefont
  {Knap}}]{RareRegions_rev}%
  \BibitemOpen
  \bibfield  {author} {\bibinfo {author} {\bibfnamefont {K.}~\bibnamefont
  {Agarwal}}, \bibinfo {author} {\bibfnamefont {E.}~\bibnamefont {Altman}},
  \bibinfo {author} {\bibfnamefont {E.}~\bibnamefont {Demler}}, \bibinfo
  {author} {\bibfnamefont {S.}~\bibnamefont {Gopalakrishnan}}, \bibinfo
  {author} {\bibfnamefont {D.~A.}\ \bibnamefont {Huse}}, \ and\ \bibinfo
  {author} {\bibfnamefont {M.}~\bibnamefont {Knap}},\ }\href {\doibase
  10.1002/andp.201600326} {\bibfield  {journal} {\bibinfo  {journal} {Ann.
  Phys.}\ }\textbf {\bibinfo {volume} {529}},\ \bibinfo {pages} {1600326}
  (\bibinfo {year} {2017})},\ \bibinfo {note} {1600326}\BibitemShut {NoStop}%
\bibitem [{\citenamefont {Ponte}\ \emph {et~al.}(2017)\citenamefont {Ponte},
  \citenamefont {Laumann}, \citenamefont {Huse},\ and\ \citenamefont
  {Chandran}}]{Thermal_inclusions}%
  \BibitemOpen
  \bibfield  {author} {\bibinfo {author} {\bibfnamefont {P.}~\bibnamefont
  {Ponte}}, \bibinfo {author} {\bibfnamefont {C.~R.}\ \bibnamefont {Laumann}},
  \bibinfo {author} {\bibfnamefont {D.~A.}\ \bibnamefont {Huse}}, \ and\
  \bibinfo {author} {\bibfnamefont {A.}~\bibnamefont {Chandran}},\ }\href@noop
  {} {\bibfield  {journal} {\bibinfo  {journal} {Phil. Trans. R. Soc. A}\
  }\textbf {\bibinfo {volume} {375}},\ \bibinfo {pages} {20160428} (\bibinfo
  {year} {2017})}\BibitemShut {NoStop}%
\bibitem [{\citenamefont {Chandran}\ \emph {et~al.}(2016)\citenamefont
  {Chandran}, \citenamefont {Pal}, \citenamefont {Laumann},\ and\ \citenamefont
  {Scardicchio}}]{Scardicchio_allD}%
  \BibitemOpen
  \bibfield  {author} {\bibinfo {author} {\bibfnamefont {A.}~\bibnamefont
  {Chandran}}, \bibinfo {author} {\bibfnamefont {A.}~\bibnamefont {Pal}},
  \bibinfo {author} {\bibfnamefont {C.~R.}\ \bibnamefont {Laumann}}, \ and\
  \bibinfo {author} {\bibfnamefont {A.}~\bibnamefont {Scardicchio}},\ }\href
  {\doibase 10.1103/PhysRevB.94.144203} {\bibfield  {journal} {\bibinfo
  {journal} {Phys. Rev. B}\ }\textbf {\bibinfo {volume} {94}},\ \bibinfo
  {pages} {144203} (\bibinfo {year} {2016})}\BibitemShut {NoStop}%
\end{thebibliography}%

\end{document}



\setcounter{equation}{0}
\setcounter{figure}{0}
\setcounter{table}{0}
\setcounter{page}{1}
\makeatletter
\renewcommand{\theequation}{S\arabic{equation}}
\renewcommand{\thefigure}{S\arabic{figure}}
\renewcommand{\thetable}{S\arabic{table}}
\renewcommand{\bibnumfmt}[1]{[S#1]}
\renewcommand{\citenumfont}[1]{S#1}

\title{Supplemental Materials for:  On the possibility of many-body localization in a long-lived finite-temperature ultracold quasi-neutral molecular plasma}

\author{John Sous}  
\affiliation{Department of Physics \& Astronomy, University of British Columbia, Vancouver, BC V6T 1Z3, Canada}
\affiliation{Stewart Blusson Quantum Matter Institute, University of British Columbia, Vancouver, British Columbia, V6T 1Z4, Canada}
\author{Edward Grant}
\email[Author to whom correspondence should be addressed. Electronic mail:  ]
{edgrant@chem.ubc.ca}
\affiliation{Department of Physics \& Astronomy, University of British Columbia, Vancouver, BC V6T 1Z3, Canada}
\affiliation{Department of Chemistry, University of British Columbia, Vancouver, BC V6T 1Z3, Canada}

\maketitle

\section{Double-resonant production of a state selected molecular Rydberg gas}

Laser pulses, $\omega_1$ and $\omega_2$, cross a molecular beam to define a Gaussian ellipsoidal volume in which a sequence of resonant electronic transitions transfer population from the X $^2\Pi_{1/2}$ ground state of nitric oxide to an intermediate state, A $^2\Sigma^+$ with angular momentum neglecting spin, $N'=0$, and then to a specified level in the mixed $n_0f(2)$ Rydberg series to create a state-selected Rydberg gas of nitric oxide, in which quantities (0) and (2) refer to rotational quantum numbers of the NO$^+$ $^1\Sigma^+$ cation core.   

The intensity of $\omega_1$ determines the density of Rydberg molecules formed by saturated absorption of $\omega_2$.  For a given $\omega_1$ intensity, the peak Rydberg gas density varies with $\omega_1 - \omega_2$ delay according to the  well-known decay rate of the A $^2\Sigma^+$ state.  Choosing $I_{\omega_1}$ and $\Delta t_{\omega_1 - \omega_2}$, we precisely control the initial peak density of the Rydberg gas ellipsoid over a two-decade range from $\rho_0=10^{10}$ to $10^{12}~{\rm cm^{-3}}$ \cite{MSW_tutorial}.  

In the core of this ellipsoid, Rydberg molecules, propagating in the molecular beam have a local longitudinal temperature of $T_{||}=500$ mK and a transverse temperature, $T_{\perp}<5$ mK.  These molecules interact at a density-determined rate to form NO$^+$ ions and free electrons.  Initially created electrons collide with Rydberg molecules to trigger electron-impact avalanche on a time-scale that varies with density from nanoseconds to microseconds (see below).  

\section{Selective field ionization spectroscopy of electron binding energy }

Selective field ionization (SFI) produces an electron signal waveform that varies with the amplitude of a linearly rising electrostatic field.  Electrons in a Rydberg state with principal quantum number, $n$, ionize diabatically when the field amplitude reaches the electron binding energy threshold, $1/9n^2$ \cite{Gallagher}.  

For low density Rydberg gases, SFI has served as an exacting probe of the coupling of electron orbital angular momentum coupling with core rotation.  Studies of nitric oxide in particular have shown that $nf(2)$ Rydberg states of NO traverse the Stark manifold to form NO$^+$ in rotational states $N^+=2$ and 0 \cite{Fielding}.  

Experiments described in the main text operate in a diabatic regime, employing a slew rate of 0.7 V cm$^{-1}$ ns$^{-1}$.  Under these conditions, SFI features that appear when the field rises to an amplitude of $F$ V cm$^{-1}$ measure electrons bound by energy $E_b$  in cm$^{-1}$, according to $E_b =4 \sqrt{F}$.  

Quasi-free electrons, weakly bound in the attractive potential of more than one cation, ionize at a low field that varies with the number of excess ions in the plasma.  

\begin{figure}[h!]
    \centering
        \includegraphics[width=.9\columnwidth]{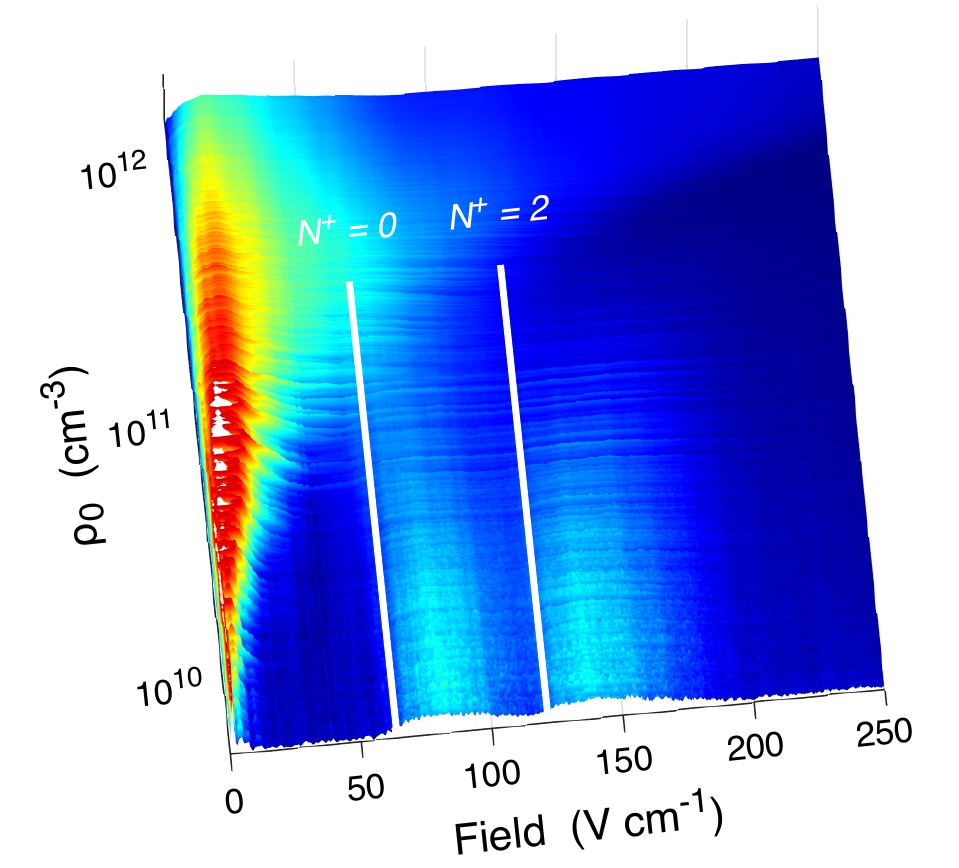}
    \caption {Selective field ionization spectrum spectrum as a function of initial Rydberg gas density, $\rho_0$, after 500 ns of evolution, showing the signal of weakly bound electrons combined with a residual population of $49f(2)$ Rydberg molecules, (initial principal quantum number, $n_0=49$, in the $f$ Rydberg series converging to NO$^+$ ion rotational state, $N^+=2$).  After 10 $\mu$s, this population sharpens to signal high-$n$ Rydbergs and plasma electrons, with a residue of the initial Rydberg population, shifted slightly to deeper binding energy by $l$-mixing and perhaps some small relaxation in $n$. The prominent feature that appears at the lowest values of the ramp field gauges the potential energy of electrons in high Rydberg states bound to single NO$^+$ ions, combined with electrons bound to the space charge of more than one ion.  Notice the binding effect of a slightly greater excess positive charge at the highest initial Rydberg gas densities.  The red feature extends approximately to the binding energy of $n_0=80$ or 500 GHz.}   \label{fig:SFI49}
\end{figure}

The SFI spectrum presented in the text as Figure 1(c) and shown here as Figure \ref{fig:SFI49} maps the electron binding energy as a function of the initial Rydberg gas density for a molecular nitric oxide ultracold plasma after 500 ns of evolution.  At a glance, the spectrum at higher density  ($10^{12}~{\rm cm}^{-3}$) shows direct evidence of either electrons bound to an increasing space charge or a broader distribution of high-$n$ Rydberg states.  

This contracts to a narrower distribution of very weakly bound electrons in plasmas of lower density ($10^{10}~{\rm cm}^{-3}$).  Here we observe the spectrum of a residue of molecules with the originally selected principal quantum number of the Rydberg gas, shifted slightly to deeper apparent binding energy by evident $l$-mixing or slight relaxation in $n$. 

We have used SFI measurements like these to characterize the avalanche and evolution dynamics of a great many Rydberg gases of varying density and initial principal quantum number.  Relaxation times vary, but all of these spectra evolve to form the same final spectrum of weakly bound electrons with traces of residual Rydberg gas for systems of low initial density.   

\section{Coupled rate-equation simulations of the electron-impact avalanche to ultracold plasma in a molecular Rydberg gas}

The semi-classical mechanics embodied in a system of coupled rate equations serves well to describe the avalanche of a molecular Rydberg gas to ultracold plasma.  In this picture, Rydberg molecule densities, labeled $\rho_i$, evolve over a ladder of principal quantum numbers, $n_i$, according to:

\begin{eqnarray}
-\frac{d\rho_i}{dt}&=&\sum_{j}{k_{ij}\rho_e\rho_i}-\sum_{j}{k_{ji}\rho_e\rho_j} \nonumber\\
&& +k_{i,ion}\rho_e\rho_i-k_{i,tbr}\rho^3_e + k_{i,PD}\rho_i
  \label{level_i}
\end{eqnarray}
\noindent The free-electron density changes as:
\begin{equation}
\frac{d\rho_e}{dt}=\sum_{i}{k_{ion}\rho_e^2}-\sum_{i}{k^i_{tbr}\rho^3_e} - k_{DR}\rho^2_e
  \label{electron}
\end{equation}

A variational reaction rate formalism determines $T_e$-dependant rate coefficients, $k_{ij}$, for electron impact transitions from Rydberg state $i$ to $j$, $k^i_{ion}$, for collisional ionization from state $i$ and $k^i_{tbr}$, for three-body recombination to state $i$ \cite{Mansbach,PVS}.  Unimolecular rate constants, $k_{i,PD}$, describe the principal quantum number dependant rate of Rydberg predissociation \cite{Bixon,GallagherNO,Remacle}, averaged over azimuthal quantum number, $l$ \cite{Chupka:1993}.  $k_{DR}$ accounts for direct dissociative recombination \cite{Schneider}

The relaxation of molecules in the manifold of  Rydberg states determines the temperature of electrons released by avalanche. Conservation of total energy per unit volume requires:
\begin{eqnarray}
E_{tot}&=&\frac{3}{2}k_BT_e(t)\rho_e(t)-R\sum_i{\frac{\rho_i(t)}{n_i^2}} \nonumber \\
&&+ \frac{3}{2}k_B T \rho_e^{DR} -R \sum_{i} \frac{\rho_i^{PD}}{n_i^2}
  \label{energy}
\end{eqnarray}
where $R$ is the Rydberg constant for NO, and $\rho_e^{DR}$ and $\rho_i^{PD}$ represent the number of electrons and Rydberg molecules of level $i$ lost to dissociative recombination and predissociation, respectively \cite{Saquet2011,Saquet2012}.

 To realistically represent the density distribution produced by crossed-beam laser excitation of the cylindrical distribution of NO ground-state molecules in the molecular beam, we use a concentric system of 100 shells of defined density spanning a Gaussian ellipsoid to 5$\sigma$ in three dimensions.  Avalanche proceeds as determined by the initial Rydberg molecule density of each shell.  Each shell conserves the combined density of stationary molecules, ions and neutral fragmentation products.  Electrons satisfy local quasi-neutrality, but are otherwise assumed mobile, and thermally equilibrated over the entire volume \cite{Haenel}.

 \begin{figure}[h!]
    \centering
        \includegraphics[width=.8 \columnwidth]{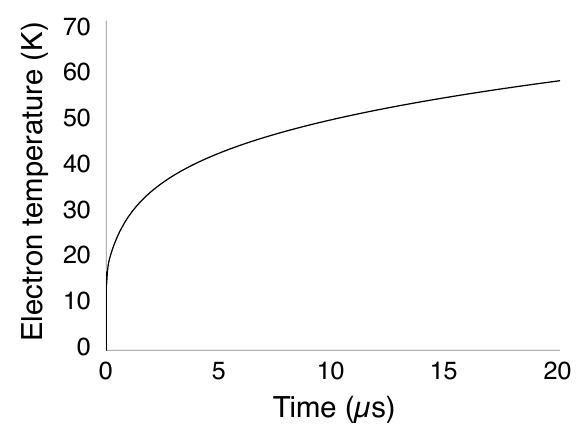}
                \includegraphics[width=.8 \columnwidth]{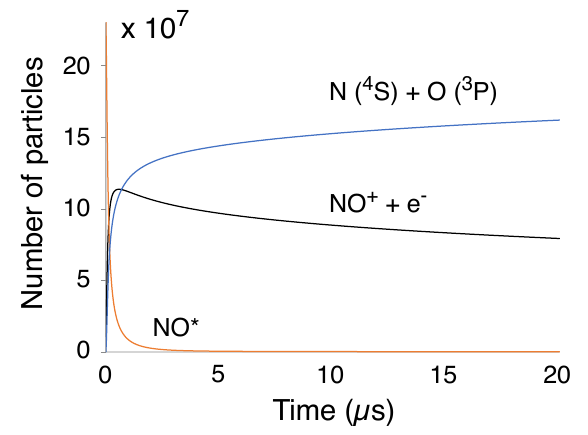}
    \caption {(lower) Numbers of ions and electrons, Rydberg molecules and neutral dissociation products N($^4$S) and O($^3$P) as a function of time during the avalanche of an $n_0=80$ Rydberg gas of NO to form an ultracold plasma, as predicted by a shell-model coupled rate equation simulation.  Here we represent the initial density distribution of the Rydberg gas by a 5$\sigma$ Gaussian ellipsoid with principal axis dimensions, $\sigma_x=1.0$ mm, $\sigma_y=0.55$ mm, $\sigma_z = 0.7$ mm and peak density of $4 \times 10^{10}~{\rm cm}^{-3}$, as measured for a typical experimental plasma entering the arrest state after am evolution of 10 $\mu$s.  The simulation proceeds in 100 concentric shells enclosing set numbers of kinetically coupled particles, linked by a common electron temperature that evolves to conserve energy globally.  (upper) Global electron temperature as a function of time.}
    \label{fig:n=80}
\end{figure}
  
  \subsection{The semi-classical evolution of an $n_0=80$ Rydberg gas}
 
 Figure \ref{fig:n=80} shows the global evolution of particle densities and electron temperature calculated for an $n_0=80$ Rydberg gas at an initial density of $4 \times 10^{10}~{\rm cm}^{-3}$ \cite{Haenel}, representing one limit of the SFI spectrum obtained as above for an ultracold plasma in its arrest state after an evolution time of 10 $\mu$s.  By this point, the real system begins a phase of unchanging composition and very slow expansion that lasts at least a millisecond -- as long a period as we can observe it.  
 
The the semi-classical simulation result shown in Figure \ref{fig:n=80} tells us that the SFI spectrum shown in Figure \ref{fig:SFI49} cannot possibly signal a conventional gas of long lived high-Rydberg molecules.  Instead, a proven semi-classical rate model configured for the density distribution of the experiment, predicts the decay of such a high-Rydberg gas to plasma on the timescale of a microsecond or less.  

In the model, predissociation consumes residual Rydbergs in all $n$-levels within a few microseconds and the formation of neutral atomic products quickly slows.  This must occur conventionally because the rising electron temperature stabilizes the classical plasma state by suppressing three-body recombination.  The real arrested state, however, shows no sign of an electron temperature higher than a few degrees K.  

  \subsection{The semi-classical evolution of a fully ionized ultracold plasma with ${T_e(0)= 5}$ K}

Let us instead test instead the kinetic stability of a conventional ultracold plasma composed entirely of ions and electrons.  Again, we assume initial conditions that fit with the observed properties of the arrest state:  NO$^+$ and electrons present at a density of $4 \times 10^{10}~{\rm cm}^{-3}$ in an ellipsoid with Gaussian dimensions, $\sigma_x=1.0$ mm, $\sigma_y=0.55$ mm, $\sigma_z = 0.7$ mm, represented by simulations evolving in 100 shells, with electron temperature equilibration \cite{Haenel}.  In keeping with the very slow rate of plasma expansion observed in the experiment, we set the initial electron temperature to 5 K.

 \begin{figure}[h!]
    \centering
        \includegraphics[width=.8 \columnwidth]{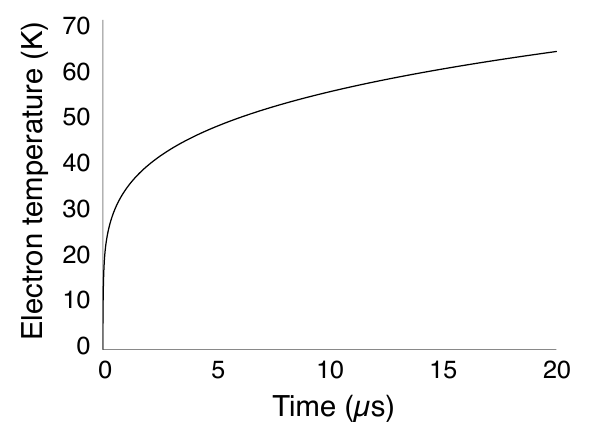}
        \includegraphics[width=.8 \columnwidth]{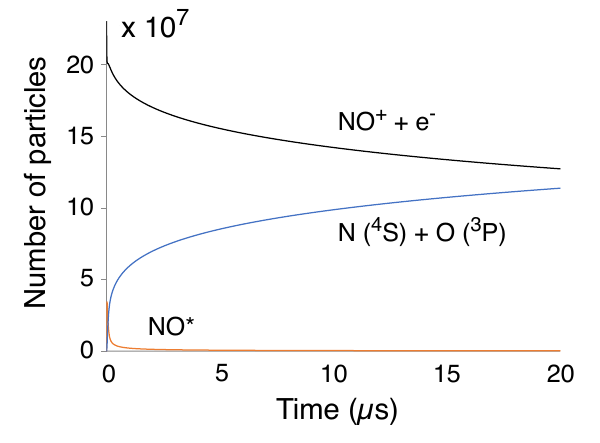}
    \caption {(lower) Numbers of ions and electrons, Rydberg molecules and neutral dissociation products N($^4$S) and O($^3$P) as a function of time during the evolution of an ultracold plasma of NO$^+$ ions and electrons, as predicted by a shell-model coupled rate equation simulation.  Here we represent the initial density distribution of the plasma by a 5$\sigma$ Gaussian ellipsoid with principal axis dimensions, $\sigma_x=1.0$ mm, $\sigma_y=0.55$ mm, $\sigma_z = 0.7$ mm, peak density of $4 \times 10^{10}~{\rm cm}^{-3}$ and initial electron temperature, $T_e(0)=5$ K, as measured for a typical experimental plasma entering the arrest state after am evolution of 10 $\mu$s.  The simulation proceeds in 100 concentric shells enclosing set numbers of kinetically coupled particles, linked by a common electron temperature that evolves to conserve energy globally.  (upper) Global electron temperature as a function of time.  }
    \label{fig:free}
\end{figure}

Figure \ref{fig:free} shows how this classical arrest state evolves in time.  The formation and rapid decay of NO Rydberg molecules signifies an immediate process of three-body recombination, which decreases the charged particle density of the plasma,  Predissociation reduces the steady-state density of Rydberg molecules to a value of nearly zero, but three-body recombination persists as shown by the rising density of neutral atom fragments.  Eventually, this process slows as the electron temperature rises.  Could this hot-electron ultracold plasma represent the end state of arrested relaxation?  Absolutely not.  As detailed in the next section, a plasma with an electron temperature of 60 K would expand to a volume larger than our experimental chamber in less than 100 $\mu$s.

\section{Ambipolar expansion in a plasma with an ellipsoidal density distribution}

The self-similar expansion of a spherical Gaussian plasma is well-described by an analytic solution of the Vlasov equations for electrons and ions with self-consistent electric fields.  For a distribution of width $\sigma$, in the limit of $T_e \gg T_i$, this solution reduces to \cite{Dorozhkina}:
\begin{equation}
e \nabla{\phi} = k_B T_e \rho^{-1} \nabla{\rho}  = - k_B T_e  \frac{r}{\sigma^2}
\label{vlasov}
\end{equation}
In essence, the thermal pressure of the electron gas produces an electrostatic force that radially accelerates the ion density distribution according to the gradient in the electrostatic potential.  In approximate terms, the expanding electrons transfer kinetic energy to the ions, accelerating the distribution to an average ballistic velocity, 
\begin{equation}
k_B T_e \approx  m_i \left < {v_i^2} \right >
\label{vlasov}
\end{equation}
The velocity varies linearly with radial distance, $\partial_t r = \gamma r$, where $\gamma$ falls with time as the distribution expands, and the electron temperature cools according to $\partial_t T_e = -2 \gamma T_e$.  

To model the ellipsoidal plasma, we represent its charge distribution by a set of concentric shells.  In this shell model, the density difference from each shell $j$ to shell $j+1$ establishes a potential gradient that determines the local electrostatic force in each principal axis direction, $k$ \cite{Sadeghi:2012}:  
\begin{align}
\frac{e}{m_i}\nabla \phi_{k,j}(t) = & \frac{\partial u_{k,j}(t)}{\partial t} \notag  \\
= & \frac{k_BT_e(t)}{m_i\rho_{j}(t)} \frac{\rho_{j}(t) - \rho_{j+1}(t)}{r_{k,j}(t) - r_{k,j+1}(t)}
  \label{dr_dt}
\end{align}
\noindent where $\rho_j(t)$ represents the density of ions in shell $j$.  

The radial coordinates of each shell evolve according to its instantaneous velocity along each axis, $u_{k,j}(t)$.
\begin{equation}
\frac{\partial r_{k,j}(t)}{\partial t}=u_{k,j}(t) = \gamma_{k,j}(t) r_{k,j}(t)
  \label{dr_dt}
\end{equation}
\noindent which in turn determines shell volume and thus its density, $ \rho_j(t)$.  The electron temperature supplies the thermal energy that drives this ambipolar expansion.  Ions accelerate and $T_e$ falls according to: 
\begin{equation}
\frac{3k_B}{2}\frac{\partial T_e(t)}{\partial t}= -\frac{m_i}{\sum_{j}{N_j}}\sum_{k,j}{N_j u_{k,j}(t)\frac{\partial u_{k,j}(t)}{\partial t}}
  \label{dr_dt}
\end{equation}

 \begin{figure}[h!]
    \centering
        \includegraphics[width= 1 \columnwidth]{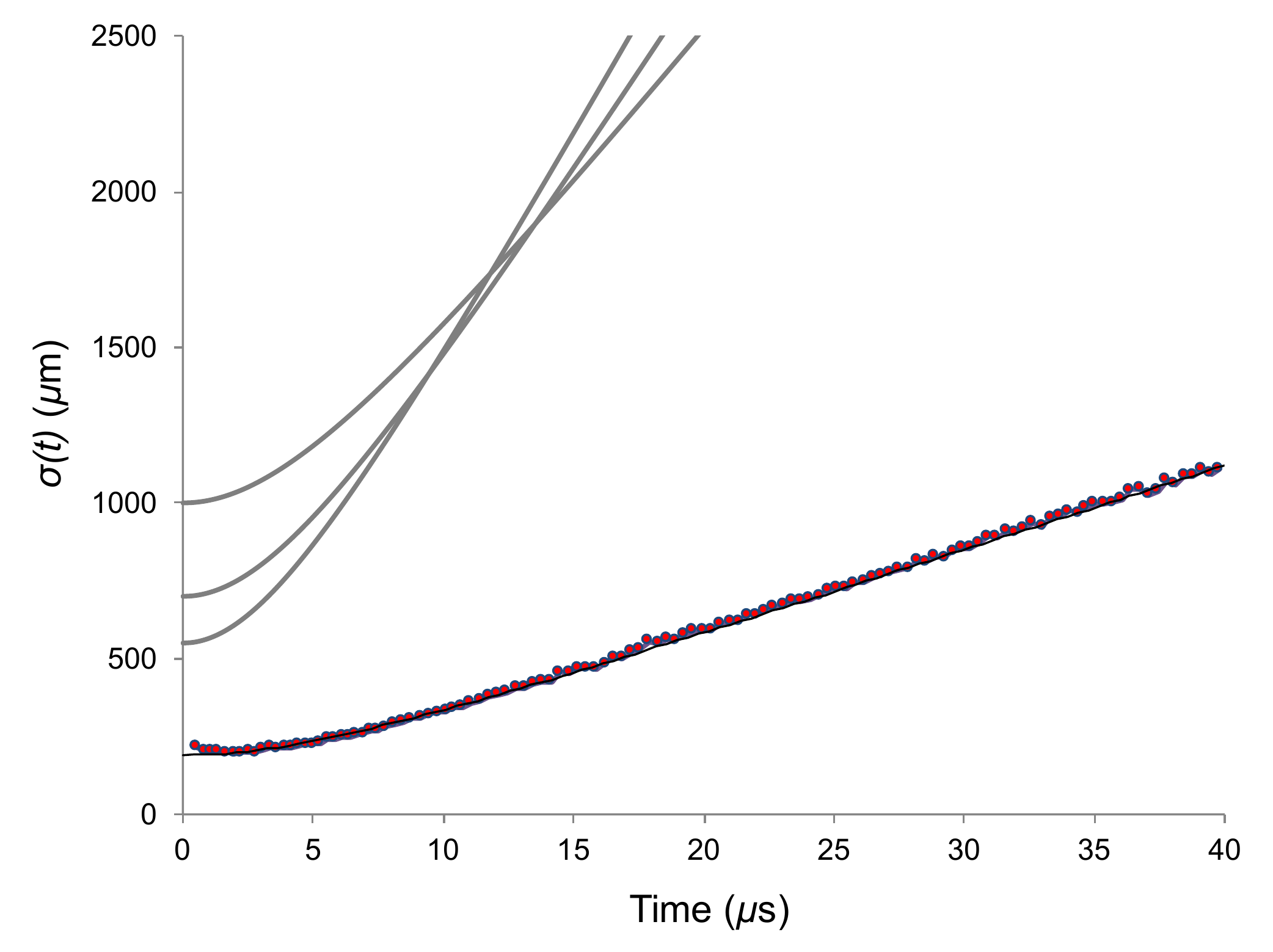}
    \caption {Hydrodynamic expansion of a Gaussian ellipsoid with the dimensions measured at 10 $\mu$s for the typical arrested plasma described above, modeled by a 100-shell simulation, assuming an electron temperature that rises to 60 K, with curves, reading from the bottom on the left, for  $\sigma_y(t)$, $\sigma_z(t)$ and $\sigma_x(t)$.  The lower curve with data shows the measured expansion of a typical molecular NO ultracold plasma with a Vlasov fit for $T_e=3$ K.  }
    \label{fig:expan}
\end{figure}

Figure \ref{fig:expan} compares the ambipolar expansion of an ellipsoidal plasma, simulated for an initial volume with the starting dimensions described above and an initial electron temperature of 60 K, compared with the time evolution of the Gaussian width measured in $z$ by experiment.  Note that the choice of a large initial volume intrinsically slows the simulated expansion.  Yet, nevertheless, the electron heating that arises inevitably from three-body recombination in a classical ultracold plasma demands a rate of expansion that is completely unsupported by experimental observation.

\section{Effective many-body Hamiltonian}
Experimental observations tell us that the molecular ultracold plasma of nitric oxide evolves to a state of arrested relaxation in which extravalent electrons occupy a narrow distribution of weakly bound states.  This distribution of states supports a vast distribution of pair-wise interactions, creating a random potential landscape. Resonant dipole-dipole interactions in this dense manifold of basis states cause excitation exchange. In the disorder potential, these processes are dominated
by low energy-excitations involving $L$ states in number, where we expect $L$ to be small (from 2 to 4). The most probable interactions select $L$-level systems composed of different basis states from dipole to dipole. Thus, the states $\ket{e^1}$, $\ket{e^2}$ ... $\ket{e^L}$ vary from one dipole to the next and from time to time. 

Representing excitations by spins, we can write an XY model \cite{Sachdevbook} that describes these interactions in terms of their effective spin dynamics 
\begin{eqnarray} \label{XY}
&H_{\rm eff}& =  \sum_i \epsilon_i \hat{S}^z_i + \sum_{i,j} J_{ij} (\hat{S}^+_i \hat{S}^-_j + h.c.)
\label{eqn:XY}
\end{eqnarray}
where $\hat{S}$ in each case denotes a spin-$L$ operator defined as $\hat{S}^\gamma = \hbar \hat{\sigma}^\gamma /2 $, for which $\sigma^\gamma$ is the corresponding spin-$L$ Pauli matrix that spans the space of the $L$ active levels and $\gamma = x,y$ or $z$. $h.c.$ refers to Hermitian conjugate.

Let us now consider specific examples of this construction.

\subsection{$L = 2$ case} 

\begin{figure}[h!]
    \centering
        \includegraphics[width= 1 \columnwidth]{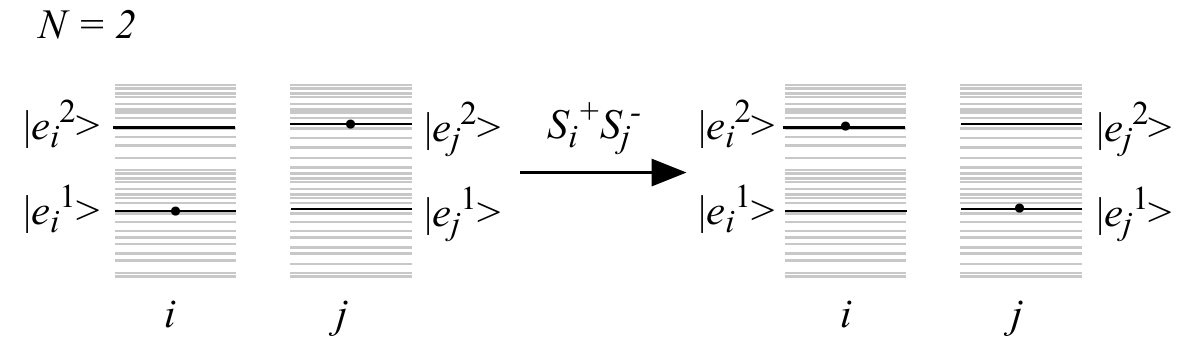}
    \caption {Schematic diagram representing two Rydberg molecules, $i$ and $j$, dipole coupled in the two-level approximation. In every case, the disorder in the environment of each molecule perturbs the exact energy level positions of $\ket{e_i}$ and $\ket{e_j}$.}
    \label{fig:N=2}
\end{figure}

Figure \ref{fig:N=2} diagrams a case that is uniquely defined for every pair of interacting dipoles.  In the limit of isolated pairs, this two-level interaction is exactly resonant.  Conditions described below randomly displace these energy level positions.  

For each particular dipole $i$, described by states $\ket{e_i^1}$ and $\ket{e_i^2}$, let us define a projection operator for the higher-energy state (which we will call spin-up) $\hat{\sigma}^{e^2}_i= \ket{e^2_i }\bra{e^2_i }= (1 + \hat{\sigma}_i^{z})/2$ and the lower-energy state (spin-down) $\hat{\sigma}^{e^1}_i = \ket{e^1_i} \bra{e^1_i} = (1 - \hat{\sigma}_i^{z})/2$. Thus, we can represent the two levels of a dipole $i$, with an energy spacing $\epsilon_i$, by a one-body operator $\epsilon_i \hat{S}^z_i = (\hbar \epsilon_i / 2) \hat{\sigma}^z_i$. This defines an energy $\pm \hbar \epsilon_i / 2$ depending on which state $\ket{e_i^2}$ or $\ket{e_i^1}$ is occupied, respectively, {\em i.e.} $\ket{e_i^2} \equiv \ket{\uparrow_i}$ and $\ket{e_i^1} \equiv \ket{\downarrow_i}$.  

The onsite energy is given by $\epsilon_{i} = E^{12}_i + D_i$ \cite{agranovich}, where $E^{12}_i$ is the energy separation between the two states $\ket{e_i^1}$ and $\ket{e_i^2}$ evaluated for the local Hamiltonian $h_i$. $h_i$ varies with the random potential landscape from one dipole to the next and thus is responsible for the diagonal disorder in the onsite term.  $D_{i} = \sum_{j \neq i} \bra{e^2_i,e^1_j}V^{dd}_{i,j}\ket{e^2_i,e^1_j} - \bra{e^1_i,e^1_j}V^{dd}_{i,j}\ket{e^1_i,e^1_j}$ represents the shift in a dipole's energy due to dipole-dipole interactions \cite{agranovich}. This term is identically zero for parity-conserving states \cite{BrownC}.

Lowering and raising operators, $\hat{\sigma}^{-}_{i} = \ket{e^1_i} \bra{e^2_i}$ and its Hermitian conjugate $\hat{\sigma}^{+}_{i} = \ket{e^2_i} \bra{e^1_i}$, define a resonant spin flip-flop between dipoles $i$ and $j$: $J_{ij} (\hat{S}^+_i \hat{S}^-_j + h.c.) = (\hbar J_{ij} / 2) (\hat{\sigma}^+_i \hat{\sigma}^-_j + h.c.)$ with amplitude $J_{ij} = {t_{ij}}/{{ r}_{ij} ^3}$; $t_{ij} = \bra{e^2_i,e^1_j}V^{dd}_{i,j}\ket{e^1_i,e^2_j}$. This refers to the dipole-dipole mediated transfer of excitation \cite{agranovich} represented by, for example, $\hat{S}^+_i \hat{S}^-_j \ket{\downarrow_i} \ket{\uparrow_j} = \ket{\uparrow_i} \ket{\downarrow_j}$ {\em i.e.} $\ket{e_i^2} \ket{e_j^1} \xrightarrow[]{\hat{S}^+_i \hat{S}^-_j} \ket{e_i^1} \ket{e_j^2}$. We can expect this class of matrix element to be non-zero for many of the local eigenstates of $h_i$ and $h_j$, as the dipole-dipole operator couples states of different parity, limited only by a few selection rules \cite{BrownC}. 

Additionally, we note that dipole-dipole interactions lead to a two-body Ising term of the form $\hat{S}^z_i \hat{S}^z_j$. This term originates from dipole-dipole induced shifts of pairs of dipoles  \cite{agranovich} and has an amplitude $\bra{e^2_i,e^2_j}V^{dd}_{i,j}\ket{e^2_i,e^2_j} + \bra{e^1_i,e^1_j}V^{dd}_{i,j}\ket{e^1_i,e^1_j}$. This term is also identically zero for parity conserving states \cite{BrownC}.  Since, the arrested phase includes no external parity-breaking fields and neglecting local field fluctuations, we assume $D_i = 0 \rightarrow \epsilon_{i} = E^{12}_i$ and no dipole-dipole induced Ising interaction.

\subsection{$L > 2$ cases}
\begin{figure}[h!]
    \centering
        \includegraphics[width= 1 \columnwidth]{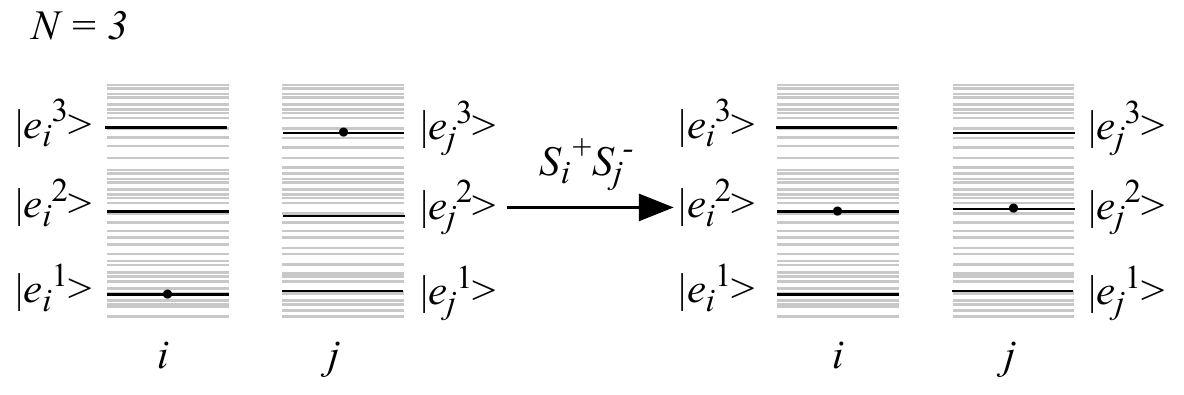}
    \caption {Schematic diagram representing two Rydberg molecules, $i$ and $j$, dipole coupled in the limits of $L=3$.  In the very high state density of the quenched ultracold plasma, the displacement of $\ket{e_i^2}$ and $\ket{e_j^2}$ will will lessen the significance of $L=3$ interactions compared with the case of $L=4$.}
    \label{fig:N=3}
\end{figure}

\begin{figure}[h!]
    \centering
        \includegraphics[width= 1 \columnwidth]{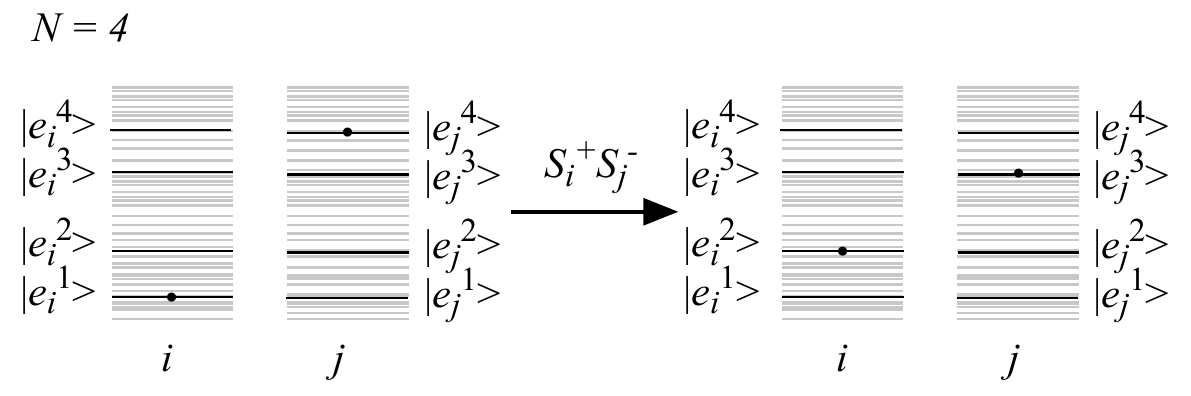}
    \caption {Schematic diagram representing two Rydberg molecules, $i$ and $j$, dipole coupled in the limits of $L=4$. The high state density and strong disorder in the quenched ultracold plasma gives this case of $L=4$ greater significance than the restrictive limit of $L=3$}
    \label{fig:N=4}
\end{figure}

We can easily imagine systematic coupling schemes that involve three or four $L$-level interactions.  Excitation transfer still governs the dynamics via terms like $J_{ij} (\hat{S}^+_i \hat{S}^-_j + h.c.)$, where the $\hat{S}$ operators live in the active $L$-dimensional subspaces. Figures \ref{fig:N=3} and \ref{fig:N=4} schematically detail examples of these interactions.  

Figure \ref{fig:N=3} represents an interaction of overwhelming importance in studies of Rydberg quantum optics.  Typically, a narrow bandwidth laser excites a resonant pair state, such as $23P_{3/2} + 23P_{3/2} \leftrightarrow 23s + 24s$ in Cs \cite{Pillet_Cs}.  Excitation transfer in this $L = 3$ case operates for example as:
\begin{equation} 
\label{two_spins}
\hat{S}^+_i \hat{S}^-_j \ket{S_i = -1} \ket{S_j = 1} = \ket{S_i = 0} \ket{S_j = 0}, 
\end{equation}
\noindent {\em i.e.} $\ket{e_i^1} \ket{e_j^3} \xrightarrow[]{\hat{S}^+_i \hat{S}^-_j} \ket{e_i^2} \ket{e_j^2}$ \\

For a gas of Rydberg molecules occupying a dense manifold of disordered states, the case of $L=3$ becomes an operationally indistinguishable special case of the more general $L=4$ interaction, which maps onto a spin of $3/2$. 

Here, we represent the interaction as an excitation transfer that operates as: 
\begin{eqnarray} 
\label{three_spins}
&\hat{S}^+_i \hat{S}^-_j \ket{S_i  =  -3/2} \ket{S_j = 3/2} \nonumber \\
 & =\ket{S_i = -1/2} \ket{S_j = 1/2}, 
\end{eqnarray}

\noindent {\em i.e.} $\ket{e_i^1} \ket{e_j^4} \xrightarrow[]{\hat{S}^+_i \hat{S}^-_j} \ket{e_i^2} \ket{e_j^3}$. \\

We can extend such sequences to higher $L$, but low-energy resonant dipole-dipole excitation exchange in the dense manifold of basis states will most prominently involve a small number of $L$-levels per dipole.

\section{Induced Van der Waals interactions}

For $|J_{ij}| << W$, sequences of interactions add Ising terms that describe a van der Waals shifts of pairs of dipoles  \cite{Burin1}.   Consider, for example, three mutually nearest-neighbour spins $i$, $j$ and $k$ in the $L$ = 2 case.  A third-order process couples spins $i$ and $j$ via spin $k$ in the following fashion: $\ket{\downarrow_i, \uparrow_j, \uparrow_k} \xrightarrow{\hat{S}^+_i \hat{S}^-_j} \ket{\uparrow_i, \downarrow_j, \uparrow_k} \xrightarrow{\hat{S}^+_j \hat{S}^-_k} \ket{\uparrow_i, \uparrow_j, \downarrow_k} \xrightarrow{\hat{S}^+_k \hat{S}^-_i} \ket{\downarrow_i, \uparrow_j, \uparrow_k}$; defining a self interaction that changes the pairwise energies of $i$, $j$.   

$U_{ij}$ is inherently random owing to the randomness in $J_{ij}$.  It is also important to note that this limit gives rise to additional perturbative processes that renormalize the local onsite fields by van der Waals terms and slightly affect the pairwise flip-flop amplitudes \cite{Zoubi1, Zoubi2, Burin1}.  We simply absorb these effects in the definitions of $\epsilon_i$ and $J_{ij}$.

Taken together with Eq (\ref{XY}) this result yields a general spin model with dipole-dipole and van der Waals interactions:
\begin{eqnarray} \label{S_XY-Ising}
H_{\rm eff}& = & \sum_i \epsilon_i \hat{S}^z_i + \sum_{i,j} J_{ij} (\hat{S}^+_i \hat{S}^-_j + h.c.) \nonumber \\
&+& \sum_{i,j} U_{ij} \hat{S}^z_i \hat{S}^z_j 
\end{eqnarray}
where $U_{ij} = {D_{ij}}/{r}_{ij}^6$ and $D_{ij} = {t_{ij}^2\widetilde{J}}/{W^2}$. 

The appearance of this third term underlines the many-body nature of Eq (\ref{XY}).  Even in this extreme limit, its dynamics are non-trivial, clearly involving more than spin flip-flops with emergent correlations between spins.

{\it Non-resonant spin-spin interactions ---} The appearance of the term, $ \sum_{i,j} U_{ij} \hat{S}^z_i \hat{S}^z_j$,  underlines the many-body nature of this model.  One obtains this term by treating $J_{ij}$ as a perturbation in Eq (\ref{eqn:XY}) \cite{Burin1}.  For the $L = 2$ case, this occurs at the third order, while for all other $L$, this term appears at the second order \cite{Burin1}. Thus, such a term arises generally in the $|J_{ij}| \ll W$ limit in three dimensions.

The van der Waals interactions occur with an amplitude, $U_{ij} \approx {J_{ij}^2\widetilde{J}}/{W^2}$, where $\widetilde{J}$  estimates $J_{ij}$ at the average distance separating spins.  We do not expect these interactions to depend strongly on the off-diagonal disorder, as they arise from the off-resonant part of $\sum_{i,j} J_{ij} (\hat{S}^+_i \hat{S}^-_j + h.c.)$, which presumably does not cause real transitions \cite{Burin1}.  Thus, we can rationalize the use of $\widetilde{J}$ here as an average weighting term.  We leave the task of studying the effect of off-diagonal disorder to future work.  

{\it Non-resonant onsite interactions ---} It is also important to note that this limit gives rise to additional perturbative processes that renormalize the local onsite fields $\sum_i \epsilon_i \hat{S}^z_i$ by van der Waals terms \cite{Burin1}.
 
Similar considerations from a completely different atomistic perspective verify that this term is approximately $ \sum_{l\neq i}{hC^{ij}_6}/{{r}_{ij}^6}$ where $h$ is the Planck constant and $C^{ij}_6$ denotes the $C_6$ coefficients for the van der Waals interaction between the off-resonant dipoles $i$ and $j$ \cite{Zoubi1, Zoubi2}. 

The induced onsite terms will also vary randomly owing to the randomness in the potential landscape.  We simply absorb such terms in the definition of $\epsilon_i$. 

\section{Resonance counting and the number of dipoles in the quenched ultracold plasma}
 
 Ref \cite{Burin1} considers the problem of delocalization via resonance counting arguments in the model of Eq \ref{S_XY-Ising} for the general case of $\alpha < \beta$, under conditions for which $d > d_c$.  Here $\alpha$ refers to the power law that regulates $J_{ij}$ and $\beta$ refers to $U_{ij}$.  $d$ and $d_c$ stand for dimensionality and critical dimensionality.  This work concludes that delocalization occurs at arbitrary disorder given sufficient system size.  
 
 For local disorder, $W$, and average spin flip-flop amplitude, $\widetilde{J}$, the resonant pair criterion defines, $N_c$, a critical number of dipoles above which the system delocalizes.  Here, we compare this theoretical estimate with an accurate experimental measure of the number of dipoles present in the arrest state of the quenched ultracold plasma. 
 
Controlled conditions of supersonic expansion precisely define the cylindrical density distribution of nitric oxide in the molecular beam \cite{MSW_tutorial}.  Co-propagating laser beams, Gaussian $\omega_1$ and $\omega_2$, cross orthogonally in the $x,y$ plane to define a Gaussian ellipsoidal excitation volume.  

When $\omega_2$ saturates the second step of double resonance, the intensity of $\omega_1$ controls the peak density of the Rydberg gas volume up to a maximum of $6 \times 10^{12}~{\rm cm}^{-3}$, obtained upon saturation of the first step.  Density varies from shot to shot, and we have developed an accurate means of classifying and binning individual SIF traces according to initial Rydberg gas peak density, as displayed in Figure \ref{fig:SFI49}.  Coupled rate simulations describing the kinetics of the avalanche of Rydberg gas to plasma confirm these estimates of peak density.

\begin{table}[h!]
 \caption{ Distribution of ions in an idealized Gaussian ellipsoid shell model of a quenched ultracold plasma of NO as it enters the arrest state with a peak density of $4 \times 10^{10}~{\rm cm}^{-3}$, $\sigma_x = 1.0$ mm, $\sigma_y = 0.55$ mm and $\sigma_z = 0.70$ mm.  At this point, the quasi-neutral plasma contains a total of  $1.9 \times 10^{8}$ NO$^+$ ions (NO Rydberg molecules).  Its average density is $1.4 \times 10^{10}~{\rm cm}^{-3}$ and the mean distance between ions is 3.32 $\mu$m.} 
 \label{tab:ellipse}
 \begin{tabular}{rlccrc}  
\toprule
  \hspace{3 pt}   Shell  &  \hspace{2 pt}  Density &   Volume   &  \hspace{2 pt} Particle &    \hspace{2 pt}  Fraction  &   $a_{ws}$  \\  
Num &   \hspace{9 pt}  cm$^{-3}$ &    cm$^{3}$  & \hspace{2 pt} Number & $\times 100$ \hspace{5 pt} &  $\mu$m \\
 \hline 
    $1$ \hspace{5 pt} & $4.0 \times 10 ^{10}$  \hspace{3 pt} & $1.8 \times 10 ^{-6}$ &  \hspace{5 pt}  $7.0 \times 10 ^{4}$ &  $0.04$ \hspace{4 pt} & $1.81$  \\ 
   $2$  \hspace{5 pt} & $3.9 \times 10 ^{10}$ & $1.1 \times 10 ^{-5}$ &  \hspace{3 pt}  $4.4 \times 10 ^{5}$ & $0.23$  \hspace{4 pt} & $1.83$  \\ 
   $3$  \hspace{5 pt} & $3.7 \times 10 ^{10}$ & $9.0 \times 10 ^{-5}$ &  \hspace{3 pt}  $3.3 \times 10 ^{6}$ & $1.75$  \hspace{4 pt} & $1.86$  \\ 
   $4$  \hspace{5 pt} & $3.3 \times 10 ^{10}$ & $2.8 \times 10 ^{-4}$ &  \hspace{3 pt}  $9.3 \times 10 ^{6}$ & $4.87$  \hspace{4 pt} & $1.93$  \\ 
   $5$  \hspace{5 pt} & $2.6 \times 10 ^{10}$ & $8.3 \times 10 ^{-4}$ &  \hspace{3 pt}  $2.2 \times 10 ^{7}$ & $11.52$  \hspace{4 pt} & $2.08$  \\ 
   $6$  \hspace{5 pt} & $2.1 \times 10 ^{10}$ & $1.2 \times 10 ^{-3}$ &   \hspace{3 pt} $2.6 \times 10 ^{7}$ & $13.40$  \hspace{4 pt} & $2.26$  \\ 
   $7$  \hspace{5 pt} & $1.5 \times 10 ^{10}$ & $1.8 \times 10 ^{-3}$ &  \hspace{3 pt}  $2.8 \times 10 ^{7}$ & $14.46$  \hspace{4 pt} & $2.49$  \\ 
   $8$  \hspace{5 pt} & $1.1 \times 10 ^{10}$ & $2.4 \times 10 ^{-3}$ &   \hspace{3 pt} $2.6 \times 10 ^{7}$ & $13.81$  \hspace{4 pt} & $2.78$  \\ 
   $9$  \hspace{5 pt} & $7.4 \times 10 ^{9}$ & $3.4 \times 10 ^{-3}$  &  \hspace{3 pt}  $2.5 \times 10 ^{7}$ & $13.28$  \hspace{4 pt} & $3.19$  \\ 
 $10$  \hspace{5 pt} & $4.3 \times 10 ^{9}$ & $5.1 \times 10 ^{-3}$ &   \hspace{3 pt} $2.2 \times 10 ^{7}$ & $11.56$  \hspace{4 pt} & $3.81$  \\ 
 $11$  \hspace{5 pt} & $2.0 \times 10 ^{9}$ & $8.3 \times 10 ^{-3}$ &   \hspace{3 pt} $1.7 \times 10 ^{7}$ & $8.85$  \hspace{4 pt} & $4.89$  \\ 
 $12$  \hspace{5 pt} & $5.6 \times 10 ^{8}$ & $1.7 \times 10 ^{-2}$ &   \hspace{3 pt} $9.4 \times 10 ^{6}$ & $4.94$  \hspace{4 pt} & $7.51$  \\ 
 $13$  \hspace{5 pt} & $7.9 \times 10 ^{7}$ & $3.1 \times 10 ^{-2}$ &   \hspace{3 pt} $2.4 \times 10 ^{6}$ & $1.27$  \hspace{4 pt} & $14.47$  \\ 
 $14$ \hspace{5 pt}  & $4.0 \times 10 ^{6}$ & $5.6 \times 10 ^{-2}$ &   \hspace{3 pt} $2.3 \times 10 ^{5}$ & $0.12$  \hspace{4 pt} & $38.96$  \\ 
 $15$  \hspace{5 pt} & $4.4 \times 10 ^{4}$ & $1.0 \times 10 ^{-1}$ &   \hspace{3 pt} $4.6 \times 10 ^{3}$ & $0.00$  \hspace{4 pt} & $176.22$  \\ 
 \hline
\end{tabular}
\end{table}

Two methods of plasma tomography determine the evolution of plasma size and relative density distribution as a function of time.  In the SFI apparatus, a perpendicular imaging grid that translates in the molecular beam propagation direction, $z$, yields an electron signal waveform that gauges the changing plasma density and width as a function of evolution time.  This waveform, followed to a point of evident arrest at about 5 $\mu$s, and well beyond, as illustrated by Figure 1 in the main text, establish a case for arrested relaxation.

Images projected in the $x,y$ plane together with waveforms in $z$, recorded after nearly 0.5 ms of flight, detail a slow ballistic expansion in Cartesian coordinates that we extrapolate back to an evolution time of 10 $\mu$s to determine the absolute density distribution of the arrested ultracold plasma, described by the shell model presented in Table \ref{tab:ellipse}.  This representation neglects the redistribution of charge density associated with the initial stages of bifurcation.  The total number of ions represented by this distribution remains constant for as long as we can measure it in our long flight-path instrument, at least a half millisecond. 

The ion density averaged over shells determines $\left <|{\bf r}_{ij}| \right > $.  This average distance between dipoles, combined with a a rough upper-limiting estimate of the average dipole-dipole matrix element, $\left < t_{ij} \right >$, based on values computed for a $\Delta n =0$ F\"oster resonant interaction in Li \cite{Zoubi2}, yields an upper-limiting estimate of $\widetilde{J}$.  

However, interaction with charged particles in the plasma environment perturbs the electronic structure of individual Rydberg molecules.  This diminishes the probability of finding resonant target states, decreasing the real value of $\widetilde{J}$, and giving rise to a rarity and randomness of resonant dipole-dipole interactions distributed over a huge state space defined by the measured distribution of electron binding energies, $W$.  

As noted in Figure \ref{fig:SFI49}, a simple measure of the width of the plasma feature in the delayed SFI spectrum determines $W$.  Table \ref{tab:summary} summarizes this and other parameters of the arrest state derived from experiment, including the $\widetilde{J}$ for Li under our conditions as an upper limit.

For short range interactions in a one-dimensional spin chain, perturbative arguments applied to disordered interacting spin models, such as the one above, predict many-body localization \cite{Imbrie}.  However, in higher dimensions especially, long-range resonant interactions play an important role in defining the conditions under which localization can occur.  It is generally accepted that interactions governed by a coupling amplitude, $J_{ij}$ that decreases with distance as $1/r_{ij}^\beta$ delocalizes any system at finite temperature for which the dimension, $d$ exceeds $\beta/2$.  

However, building on ideas introduced by Anderson \cite{Anderson} and Levitov \cite{Levitov}, Burin \cite{Burin1} offers a means by which to test a dimensionally constrained system for conditions that favor the onset of delocalization.  He uses a perturbation approach that defines limits over which localization can occur in a system as modeled above in which delocalization proceeds by the Ising interaction of extended resonant pairs.  

In this picture, a system that violates the dimension constraint delocalizes for an arbitrary size of disorder whenever the number of dipoles exceeds a critical number, $N_c$, which is determined by the disorder width, $W$ and the average coupling strength, $\tilde J$.  Coupling terms in the Hamiltonian defined by Eq \ref{S_XY-Ising} scale in ${r}$ according to $\alpha = 3$, $\beta = 6$ and $d = 3$.  This sets a critical number of dipoles defined by the quantity, $N_c = (W/\widetilde{J})^4$  \cite{Burin1}.  For the arrest state defined by the density distribution described by the elliptical shell model in Table \ref{tab:ellipse}, the measured $W$ taken with our upper limiting estimate for $\widetilde{J}$, yields $N_c = 3.6 \times 10^9$.

\begin{table}[h!]
 \caption{Resonance counting parameters in the arrest state of the quenched ultracold plasma.  The disorder width $W$, taken directly from the width of the plasma feature in the SIF spectrum, combined with $\widetilde{J}$ -- derived from a rough upper-limiting estimate of the average dipole-dipole matrix element, $\left < t_{ij} \right >$, based on values computed for $\Delta n =0$ interactions in alkali metals \cite{Zoubi2}, together with the mean distance between NO$^+$ ions in the shell model ellipsoid -- determines $N_c$. a critical number of dipoles required for delocalization.  $R^*$ describes the length scale for delocalization and $\tau^*$ denotes the delocalization time, given a sufficient number of dipoles at the average density of the experiment.  Note that the ultracold plasma quenched experimentally contains an order of magnitude fewer than $N_c$ dipoles.  } 
  \label{tab:summary}
 \begin{tabular}{ccccccc}  
 \toprule 
$W$   & $ \left < t_{ij} \right >$ &    \hspace{1 pt}   $\left <|{\bf r}_{ij}| \right > $     \hspace{1 pt}  &   $\widetilde{J}$  &   $N_c$   &  $R^*$   &     $\tau ^*$  \\ 
GHz      & \hspace{1 pt} GHz($\mu$m)$^3$   \hspace{1 pt}  &   $\mu$m   &    GHz     &       &   $\mu$m     &    s   \\
 \hline
 500  &  75  &  3.3  &  2.0  &  \hspace{3 pt} $3.6 \times 10^9$  \hspace{3 pt} &   4000  &  0.85  \\
 \hline 
\end{tabular}
\end{table}

Considering this value of $N_c$ in relation to the average density of the system at arrest defines $R^*$, an effective distance between resonant dipoles at which point this occurs \cite{Burin1}.  For the conditions described in Table \ref{tab:ellipse}, this coupling would occur in a system large enough to contain $3.6 \times 10^9$ dipoles a distance of 4 mm or more.  At this distance, our upper-limiting dipole-dipole matrix element would predict a characteristic irreversible transition time, $\tau^*$ on the order of one second \cite{Burin2}.  
 
We note that the quenched ultracold plasma formed experimentally relaxes to a volume that contains an order of magnitude fewer dipoles than $N_c$, as determined for this case by the model of Ref  \cite{Burin1}.   

As we attempt to convey above, the experiment yields plasmas of well defined density distribution and total number of dipoles.  However, the precise nature of the associated quantum states and their dipole-dipole interaction is much less well known.  This limits the certainty with which we can determine $N_c$.  What's more perturbation theory in a locator expansion formulation may not accurately define the limiting conditions for MBL in higher dimensions \cite{Sondhi}.  

Rare thermal regions (Griffiths regions) are thought to destabilize MBL systems of higher dimension \cite{Sarang1, Sarang2, Roeck_griffith, RareRegions_rev, Thermal_inclusions}, creating a glassy state, characterized by a slow evolution to a delocalized phase.  However, other results contradict this notion, and support the possibility of localization in all dimensions \cite{Scardicchio_allD}.  An added feature in the self-assembly of the molecular ultracold plasma may preclude destabilization by rare thermal regions:  Should the quenched plasma develop a Griffiths region as a site for delocalization to occur, the predissociation of relaxing NO molecules would promptly deplete that region to a void of no consequence.  
 
In any event, the quenched plasma seems consistently able to find the conditions necessary for arrested relaxation.  A great many different avalanche starting conditions, defined by varying initial Rydberg gas density and initial principal quantum number, all evolve to retain comparable internal energy and yield an arrest state with much the same density distribution as that described by the shell model detailed in Table \ref{tab:ellipse}.  \\

\begin{acknowledgements}

This work was supported by the US Air Force Office of Scientific Research (Grant No. FA9550-12-1-0239), together with the Natural Sciences and Engineering research Council of Canada (NSERC), the Canada Foundation for Innovation (CFI), the British Columbia Knowledge Development Fund (BCKDF) and the Stewart Blusson Quantum Matter Institute (SBQMI).  JS gratefully acknowledges support from the Harvard-Smithsonian Institute for Theoretical Atomic, Molecular and Optical Physics (ITAMP).  We have benefited from helpful interactions with Rahul Nandkishore, Shivaji Sondhi and Alexander Burin.  We also gratefully acknowledge discussions with Rafael Hanel, James Keller, Rodrigo Vargas, Arthur Christianen, Markus Schulz-Weiling, Hossein Sadeghi and Luke Melo.  

\end{acknowledgements}

\bibliography{Rydberg_localization_supp}